\documentclass[aps,prb,twocolumn,floatfix,superscriptaddress,showpacs]{revtex4-1}
\pdfoutput=1
\usepackage{epsfig}
\usepackage{amsmath}
\usepackage{amssymb}
\usepackage{wasysym}
\usepackage{MnSymbol}
\usepackage[colorlinks=true,citecolor=blue]{hyperref}

\newcommand{\bs}{\boldsymbol}

\begin{document}

\title{A comparative DMFT study of the $e_g$-orbital Hubbard model in thin films}

\author{Andreas R\"uegg}
\affiliation{Department of Physics, University of California, Berkeley, CA 94720, USA}
\affiliation{Theoretische Physik, Wolfgang-Pauli-Strasse 27, ETH Z\"urich, CH-8093 Z\"urich, Switzerland }
\author{Hsiang-Hsuan Hung}
\affiliation{Department of Physics, The University of Texas at Austin, Austin, TX 78712, USA}
\author{Emanuel Gull}
\affiliation{Department of Physics, University of Michigan, Ann Arbor, MI 48109, USA}
\affiliation{Max-Plank Institute for Complex Systems, Dresden, Germany}
\author{Gregory A.~Fiete}
\affiliation{Department of Physics, The University of Texas at Austin, Austin, TX 78712, USA}

\begin{abstract}
Heterostructures of transition-metal oxides emerged as a new route to engineer electronic systems with desired functionalities. Motivated by these developments, we study a two-orbital Hubbard model in a thin-film geometry confined along the cubic [001] direction using the dynamical mean-field theory. We contrast the results of two approximate impurity solvers (exact diagonalization and one-crossing approximation) to the results of the numerically exact continuous-time quantum Monte Carlo solver. Consistent with earlier studies, we find that the one-crossing approximation performs well in the insulating regime, while the advantage of the exact-diagonalization based solver is more pronounced in the metallic regime. We then investigate various aspects of strongly correlated $e_g$-orbital systems in thin film geometries. In particular, we show how the interfacial orbital polarization dies off quickly a few layers from the interface and how the film thickness affects the location of the interaction-driven Mott transition. In addition, we explore the changes in the electronic structure with varying carrier concentration and identify large variations of the orbital polarization in the strongly correlated regime. 
\end{abstract}
\date{\today}
\pacs{71.10.-w, 71.27.+a, 73.40.-c}
\maketitle
\section{Introduction}

Correlated oxide heterostructures have emerged as a new experimental path to obtain and control electronic states with unusual properties at interfaces, in quantum wells or superlattices. A broad spectrum of physical phenomena has been realized already in artificial structures, including insulator-metal transitions, superconductivity, magnetism, ferroelectricity, multi-ferroic behavior as well as integer and fractional quantum Hall phases.\cite{Mannhart:2008,Zubko:2011,Hwang:2012,Chakhalian:2012} This rich behavior is commonly attributed to a delicate interplay between spin, charge, orbital and lattice degrees of freedom, making the electronic properties of transition-metal oxides rather susceptible to the presence of local symmetry breaking, charge transfer or strain introduced by the heterostructuring.

The wide range of possibilities to combine different materials with comparable structural properties and the prospect to grow structures with atomic precision also stimulated more ambitious theoretical proposals. These include the idea to mimic the electronic structure of cuprate high-temperature superconductors in nickelate heterostructures,\cite{Chaloupka:2008,Hansmann:2009} suggestions to realize topological insulator phases in sandwich structures grown along the [111] direction\cite{Xiao:2011,Yang:2011,Ruegg:2011c,Hu:2012a} or Majorana chains in one-dimensional channels\cite{Cen:2008} at the LaAlO$_3$/SrTiO$_3$ interface.\cite{Fidkowski:2013} A necessary step towards control of the electronic phases in heterostructures involves the manipulation of the orbital degrees of freedom, in particular the occupation of the $d$-orbitals of the transition-metal ions. The orbital occupations are particularly susceptible to the local symmetry-breaking and the effect of tensile or compressive strain near the interface.\cite{Okamoto:2006,May:2010,Rondinelli:2010} In addition also electronic correlations can rearrange the orbital occupation of the $d$-electrons\cite{Hansmann:2009,Hansmann:2010,Han:2011} and simultaneously accounting for all these effects  is a challenging theoretical task.

Our study is motivated by the recent interest in rare earth nickelate superlattices grown along the [001] direction of the (pseudo-)cubic perovskite unit cell.\cite{Chaloupka:2008,Hansmann:2009,Hansmann:2010,Han:2010,Chakhalian:2011,Benckiser:2011,Freeland:2011,Liu:2011,Han:2011,LeeS:2011a,LeeS:2011b,Boris:2011,Wu:2013} To make connection with these systems, we focus on a two-orbital Hubbard model for the $e_g$ orbital manifold of the $d$-shell in a thin film geometry. Due to the confinement along the $z$-direction, the degeneracy between the $d_{3z^2-r^2}$ and $d_{x^2-y^2}$ orbitals is lifted in the thin-film geometry which induces a finite orbital polarization. Using single-site dynamical-mean field theory (DMFT),\cite{Georges:1996,Kotliar:2006,Metzner:1989} we investigate the dependence of the orbital polarization on the interaction energy, the carrier concentration and the number of layers. We show that the orbital polarization is an interface effect which vanishes within about 3 layers from the interface. We also demonstrate that the orbital polarization depends rather strongly on the total carrier concentration and can change sign several times for large electron-electron interaction as function of the $d$-level occupation.

Although the $e_g$-orbitals in bulk rare earth nickelates are quarter filled (i.e.~there is an average of one electron per site), there is a two-fold motivation to study the effect of variable carrier concentration. First, it is well-known that the carrier concentration can be different at interfaces as compared to the bulk. One prominent effect is the charge transfer across the interface in polar/non-polar heterostructures in order to avoid a huge electrostatic energy from forming an electric dipole.\cite{Ohtomo:2002,Okamoto:2004a,Okamoto:2004c,Nakagawa:2006}
The second interest in the dependence on the carrier concentration comes from the ``self-doping" effect\cite{Mizokawa:2000} believed to be important for the physics of the nickelates. In fact, the actual number of electrons in the Ni $d$-shell can considerably differ from the naive ionic picture\cite{Wang:2012} due to charge-transfer from the oxygens to the Ni-ions. Including this effect by explicitly keeping oxygen states in an effective lattice model for nickelates\cite{Wang:2012,Parragh:2013} allows for an explanation of the reduction of the orbital polarization found in (001) superlattices\cite{Freeland:2011,Han:2011} and also allows for a scenario of the paramagnetic insulator state found in many nickelates.\cite{Freeland:2011,Mizokawa:2000,Park:2012}

From a methodical point-of-view, the purpose of this work is to address the quality of two approximate impurity solvers in the context of multi-orbital models in a thin film geometry. The first solver is based on a self-consistent hybridization expansion within the one-crossing approximation (OCA).\cite{Pruschke:1989,Ruegg:2013b} The second solver uses exact diagonalization (ED) \cite{Gaffarel:1994,Bolech:2003,Liebsch:2012,Zgid:2012} for a discretized bath. We benchmark both solvers against each other and against numerically exact continuous-time quantum Monte Carlo (CT-QMC).\cite{Gull:2011} Because CT-QMC methods are computationally intensive (which often prohibits surveys of large parameter spaces), and in addition may suffer from a sever ``sign problem" in certain cases, it is desirable to identify numerically cheap solvers which are reasonably accurate. While approximate solvers may perform well in some situations, they can fail in other cases and it is therefore important to test them in the physically relevant context, such as the multi-orbital models in the thin film geometry of interest in this article. Consistent with earlier studies,\cite{Gull:2010,Haule:2010} we find that the DMFT(OCA) scheme is accurate and efficient in the insulating regime for large interactions. Unfortunately, we also find that OCA is rather inaccurate in the metallic regime (even for large interactions), in contrast to what has been found in other multi-orbital systems.\cite{Haule:2010} On the other hand, DMFT(ED) is reliable both in the metallic and insulating regime but we find it more efficient in the metallic regime.

This paper is organized as follows: in Sec.~\ref{sec:Model-Method}, we introduce the layered two-orbital Hubbard model which we study within the layer-DMFT framework in the remainder of this paper. Section~\ref{sec:benchmarking} benchmarks our approximate solvers to CT-QMC results using the model introduced in Ref.~\onlinecite{Hansmann:2010}. In Sec.~\ref{sec:results}, we then use these solvers to investigate various aspects of the two-orbital Hubbard model in thin film geometries. We discuss and summarize our results in Sec.~\ref{sec:conclusions}.

\section{Model and method}
\label{sec:Model-Method}

\subsection{Two-orbital Hubbard model}
The Hubbard-type model considered in the following is motivated by experiments involving a controlled number $L$ of atomic (001) layers of LaNiO$_3$, separated by regions of LaAlO$_3$.\cite{Boris:2011} Because LaAlO$_3$ has a large band gap, it is a good approximation to replace it by vacuum; hence, we only focus on the atomic LaNiO$_3$ layers.\cite{Hansmann:2009,Hansmann:2010,Han:2011} In order to model the conduction electrons, we consider a simple two-orbital Hubbard model of $e_g$-electrons placed at the sites of the Ni-ions. Hopping is mediated through oxygen $p$-states which are located in-between the Ni sites. Under the assumption that the energy of the $p$-orbital levels is sufficiently far away from the $d$-orbital levels of the Ni$^{3+}$, the only role of the oxygen states is to induce effective hoppings between the Ni sites. In the present paper, we will work under this assumption and therefore focus on the following lattice Hamiltonian
\begin{equation}
H=H_{\rm kin}+H_{\rm cf}+H_{\rm int}.
\label{eq:H}
\end{equation}
The individual terms are specified in the following.
\subsubsection{Kinetic energy}
The kinetic energy is of the form
\begin{equation}
H_{\rm kin}=\sum_{\bs k}\sum_{i,j}d_{i{\bs k}\sigma}^{\dag}[\hat{\mathcal{E}}({\bs k})]_{ij}d_{j{\bs k}\sigma}.
\label{eq:Hkin}
\end{equation}
Here, $i$ and $j$ both denote a pair of orbital and layer index, i.e.~$i\equiv(\alpha,l)$. In the mixed representation of Eq.~\eqref{eq:Hkin}, $d_{\alpha\sigma l}^{\dag}({\bs k})$ denotes the creation operator of an $e_g$-electron in layer $l$, in orbital $\alpha=d_{3z^2-r^2}$ or $d_{x^2-y^2}$, with spin $\sigma=\uparrow$, $\downarrow$ and two-dimensional momentum ${\bs k}$.
The Bloch matrix $\hat{\mathcal{E}}({\bs k})$ is a $2L\times 2L$ matrix of the form
\begin{equation}
\hat{\mathcal{E}}({\bs k})=\begin{pmatrix}
\mathcal{E}_{xy}({\bs k})&-t_z&0&\dots&\dots\\
-t_z&\mathcal{E}_{xy}({\bs k})&-t_z&0&\dots\\
0&-t_z&\mathcal{E}_{xy}({\bs k})&-t_z&\dots\\
\vdots&\vdots&\vdots&\ddots&\vdots\\
0&0&0&-t_z&\mathcal{E}_{xy}({\bs k})
\end{pmatrix}.
\label{eq:HBloch}
\end{equation}
The growth direction of the heterostructure is denoted by $z$ and the off-diagonal blocks of $\hat{\mathcal{E}}({\bs k})$ describe the coupling between neighboring layers. In the basis $(d_{3z^2-r^2},d_{x^2-y^2})$ the interlayer hopping takes the form
\begin{equation}
t_z=t
\begin{pmatrix}
1&0\\
0&0
\end{pmatrix}
\label{eq:tz}
\end{equation}
Equation~\eqref{eq:tz} is a manifestation of the fact that nearest-neighbor hopping between different layers occurs via the $d_{3z^2-r^2}$ orbitals.
Analogous, nearest-neighbor hopping along the $x$ direction is possible between $d_{3x^2-r^2}=-\frac{1}{2}d_{3z^2-r^2}+\frac{\sqrt{3}}{2}d_{x^2-y^2}$ and along the $y$-direction between $d_{3y^2-r^2}=-\frac{1}{2}d_{3z^2-r^2}-\frac{\sqrt{3}}{2}d_{x^2-y^2}$ orbitals:
\begin{equation}
t_x=-\frac{t}{4}
\begin{pmatrix}
1&-\sqrt{3}\\
-\sqrt{3}&3
\end{pmatrix},\quad
t_y=-\frac{t}{4}
\begin{pmatrix}
1&\sqrt{3}\\
\sqrt{3}&3
\end{pmatrix}.
\end{equation}
Hence, the contribution from nearest-neighbor hopping to the in-plane Bloch matrix $\mathcal{E}_{xy}({\bs k})$ is given by 
\begin{equation}
\mathcal{E}_{xy}^{(1)}({\bs k})=
\begin{pmatrix}
-\frac{t_{11}}{2}\left(\cos k_x+\cos k_y\right)&\frac{\sqrt{3}t_{12}}{2}\left(\cos k_x-\cos k_y\right)\\
\frac{\sqrt{3}t_{21}}{2}\left(\cos k_x-\cos k_y\right)&-\frac{3t_{22}}{2}\left(\cos k_x+\cos k_y\right)
\end{pmatrix},
\label{eq:Hxy}
\end{equation}
where $t_{11}=t_{22}=t_{12}=t_{21}=t$ in the ideal system. However, to facilitate comparison with previous calculations,\cite{Hansmann:2010} we also consider the case where these amplitudes slightly differ from each other. Furthermore, we also consider a second-neighbor hopping within the layers
\begin{equation}
\mathcal{E}_{xy}^{(2)}({\bs k})=
\begin{pmatrix}
-2t'\cos k_x\cos k_y&0\\
0&-6t'\cos k_x\cos k_y
\end{pmatrix}.
\label{eq:Exy2}
\end{equation}
The total in-plane Bloch matrix entering Eq.~\eqref{eq:HBloch} is then given by
\begin{equation}
\mathcal{E}_{xy}({\bs k})=\mathcal{E}_{xy}^{(1)}({\bs k})+\mathcal{E}_{xy}^{(2)}({\bs k}).
\end{equation}
A good tight-binding fit of the band-structure of bulk LaNiO$_3$ can be obtained with a first-neighbor hopping $t\approx0.6$ eV and $t'\approx0.06$ eV.\cite{Hansmann:2010} For the relaxed heterostructures, the tight-binding parameters can slightly differ\cite{Hirayama:2012,ZhongEPL:2012,Ruegg:2012b,Ruegg:2013c} but for our purpose, the bulk parameters provide a sufficiently good estimate. Because $t'/t\approx 1/10$, we later on make a further simplification and set $t'=0$.
Finally, we also considered the bulk Hamiltonian where the Bloch matrix Eq.~\eqref{eq:HBloch} is replaced by
\begin{eqnarray}
&&\mathcal{E}_{3D}({\bs k})=\\
&&\begin{pmatrix}
-\frac{t}{2}\left(\cos k_x+\cos k_y\right)-2t\cos k_z&\frac{\sqrt{3}t}{2}\left(\cos k_x-\cos k_y\right)\\
\frac{\sqrt{3}t}{2}\left(\cos k_x-\cos k_y\right)&-\frac{3t}{2}\left(\cos k_x+\cos k_y\right)
\end{pmatrix}.\nonumber
\end{eqnarray}
\subsubsection{Crystal field splitting}
The reduction of the cubic symmetry in the thin film geometry also allows for an explicit crystal field splitting
\begin{equation}
H_{\rm cf}=\frac{\Delta}{2}\sum_{{\bs r},\sigma}\left(n_{{\bs r},{3z^2-r^2},\sigma}-n_{{\bs r},{x^2-y^2},\sigma}\right).
\label{eq:Hcf}
\end{equation}
The energy splitting $\Delta$ in general depends on the amount of strain present in the system and could in principle be layer dependent. For the single-layer model, by fitting with band-structure calculations for the LaAlO$_3$/LaNiO$_3$ heterostructure, it was found that $\Delta=0.15$ eV.\cite{Hansmann:2010} However, we remark that even if $\Delta=0$, there is an implicit crystal field in the thin-film geometry which arises from the asymmetry in the hopping and which, in the non-interacting model, lowers the energy of the $d_{x^2-y^2}$ orbital with respect to $d_{3z^2-r^2}$. For $\Delta=0.15$ eV and $n=1$, we find that the implicit crystal field is dominant and apart from Sec.~\ref{sec:benchmarking}, we therefore set $\Delta=0$. This is consistent with the detailed study on the dependence of the orbital polarization on the value of $\Delta$ recently reported in Ref.~\onlinecite{Parragh:2013}.

\subsubsection{Electron-electron interaction}
The electron-electron interaction is incorporated in a local multi-orbital interaction of the standard type 
\begin{eqnarray}
H_{\rm int}=\sum_{\bs r}\Big[U\sum_ {\alpha}n_{{\bs r}\alpha\uparrow}n_{{\bs r}\alpha\downarrow}
+(U'-J)\sum_{\alpha>\beta,\sigma}n_{{\bs r}\alpha\sigma}n_{{\bs r}\beta\sigma}\nonumber\\
+U'\sum_{\alpha\neq \beta}n_{{\bs r}\alpha\uparrow}n_{{\bs r}\beta\downarrow}\Big].
\label{eq:Hint}
\end{eqnarray}
Here, $J$ is the Hund's coupling, $U$ denotes the intra-orbital and $U'$ the inter-orbital repulsion. We use $U=U'+2J$ which is expected to be approximately fulfilled in the considered system. The Hund's rule coupling $J$ is typically of the order of $0.5$ - $1$ eV which in magnitude is similar to the nearest-neighbor hopping for our system. The intra-orbital repulsion $U$ for thin-film nickelates is less-well known. Typical values range from 5 to 7 eV which amounts to roughly $U=8t$ - $12t$. Because of this uncertainty, we consider a range of $U$ values and study how physical quantities depend on it. 

The restriction in Eq.~\eqref{eq:Hint} to only density-density rather than the full SU(2) symmetric interaction reduces the computational complexity and allows us to benchmark our calculations to previous work.\cite{Hansmann:2010} We note, however, that all the used impurity solvers are in principle capable to deal with the more complicated rotationally invariant form of the local interaction. In particular, there are several elegant ways to reduce the size of the matrix blocks of the interaction Hamiltonian, which can be used, e.g., both in OCA or hybridization-expansion CT-QMC calculations.\cite{Werner:2006b,Lauchli:2009,Parragh:2012}

\subsection{Layer-DMFT approximation}
We study the two-orbital Hubbard model within the layer-DMFT framework ($l$-DMFT) which is a straightforward generalization of the single-site DMFT equations to models which are non-uniform along one spatial direction, e.g.~the growth direction in oxide-heterostructures.\cite{Potthoff:1999,Okamoto:2004b,Okamoto:2006,Freericks:2006,Ruegg:2007,Ishida:2008} The fundamental object in this approach is the electronic self-energy $\hat{\Sigma}(i\omega_n,{\bs k})$. For a fixed Matsubara frequency $\omega_n$ and fixed two-dimensional momentum ${\bs k}$, the self-energy is a $(N L)\times (N L)$ matrix. Here, $N$ denotes the number of spin/orbital degrees of freedom and $L$ the number of layers. The central approximation in the $l$-DMFT is the assumption of a local self-energy, i.e.
\begin{equation}
\left[\hat{\Sigma}^{l-\rm DMFT}(i\omega_n,{\bs k})\right]_{jl}=\delta_{jl}\Sigma_l(i\omega_n)
\end{equation}
where $j$ and $l$ are layer indices. The diagonal entries $\Sigma_l(i\omega_n)$ are then computed from a set of multi-orbital Anderson-impurity models with layer-dependent hybridization functions $\Delta_l(i\omega_n)$. The hybridization function $\hat{\Delta}(i\omega_n)=\delta_{ll'}\Delta_l(i\omega_n)$ and self-energy satisfy the self-consistency condition
\begin{eqnarray}
\left[i\omega_n+\mu-\hat{\Delta}(i\omega_n)-\hat{\Sigma}(i\omega_n)\right]^{-1}\nonumber\\
=\frac{1}{N_s}\sum_{\bs k}\left[i\omega_n+\mu-\hat{\mathcal{E}}({\bs k})-\hat{\Sigma}(i\omega)\right]^{-1}.
\label{eq:DMFT}
\end{eqnarray}
In the limit of a single layer, Eq.~\eqref{eq:DMFT} reduces to the self-consistency of the single-site DMFT approximation.

There are different algorithms available to approach the considered layer-DMFT problem and previous studies on closely related systems used the Hirsch-Fye quantum Monte Carlo algorithm (HF-QMC)\cite{Hansmann:2009,Hansmann:2010} and the CT-QMC.\cite{Gull:2011,Han:2011} Here, we use two approximate methods to solve the two-orbital impurity model. The first solver is based on the self-consistent hybridzation expansion in the one-crossing approximation (OCA).\cite{Haule:2010,Gull:2010,Ruegg:2013b} The second one is based on exact diagonalization of a finite system (ED).\cite{Gaffarel:1994,Georges:1996,Zgid:2012,Liebsch:2012} The DMFT(ED) results were obtained by including 5 bath sites per orbital. The ground-state of the equivalent spinfull 12-site Hubbard model was obtained using the Lanczos algorithm. To solve the self-consistency relation Eq.~\eqref{eq:DMFT}, we used a fictive temperature $T_0=0.0005t$. In the following, we benchmark our DMFT(OCA) and DMFT(ED) against DMFT(CT-QMC) based on the CT-QMC code\cite{Werner:2006,Gull:2011c} available from the ALPS library.\cite{Bauer:2011} Throughout this work, we focus on the paramagnetic phases.
\section{Benchmarking}
\label{sec:benchmarking}
\subsection{Comparison with QMC}
To address the quality of our approximate solvers in the context of the layered $e_g$-Hubbard model, we first benchmark the DMFT(OCA/ED) against DMFT(CT-QMC) for model parameters specified in Ref.~\onlinecite{Hansmann:2010}. In addition to the results of our computations, we also include results published in Ref.~[\onlinecite{Hansmann:2010}], which were obtained using the DMFT(HF-QMC) scheme. We consider a single-layer $L=1$ and the microscopic parameters are fixed as follows: the nearest-neighbor hopping amplitudes in Eq.~\eqref{eq:Hxy} are $t_{11}=0.68$~eV, $t_{22}=0.6$~eV and $t_{12}=t_{21}=0.65$~eV, the second-neighbor hopping amplitude in Eq.~\eqref{eq:Exy2} is $t'=0.06$~eV and the crystal-field splitting in Eq.~\eqref{eq:Hcf} is $\Delta=0.15$~eV. The Hund's rule coupling is fixed at $J=0.7$~eV and for the intra-orbital repulsion $U$, we consider two different values: $U=4.4$ eV and $U=7.4$~eV. The inverse temperature for the OCA/QMC was fixed at $\beta =10$~eV$^{-1}$. 

Our comparison focuses on the electronic self-energy in Matsubara frequency space $\Sigma(i\omega_n)$. Within the DMFT frame-work, $\Sigma(i\omega_n)$ determines all the single-particle properties of the model. It is also sensitive to approximations made in solving the impurity problem and therefore is a suitable quantity for benchmarking.

\begin{figure}
\includegraphics[width=0.49\linewidth]{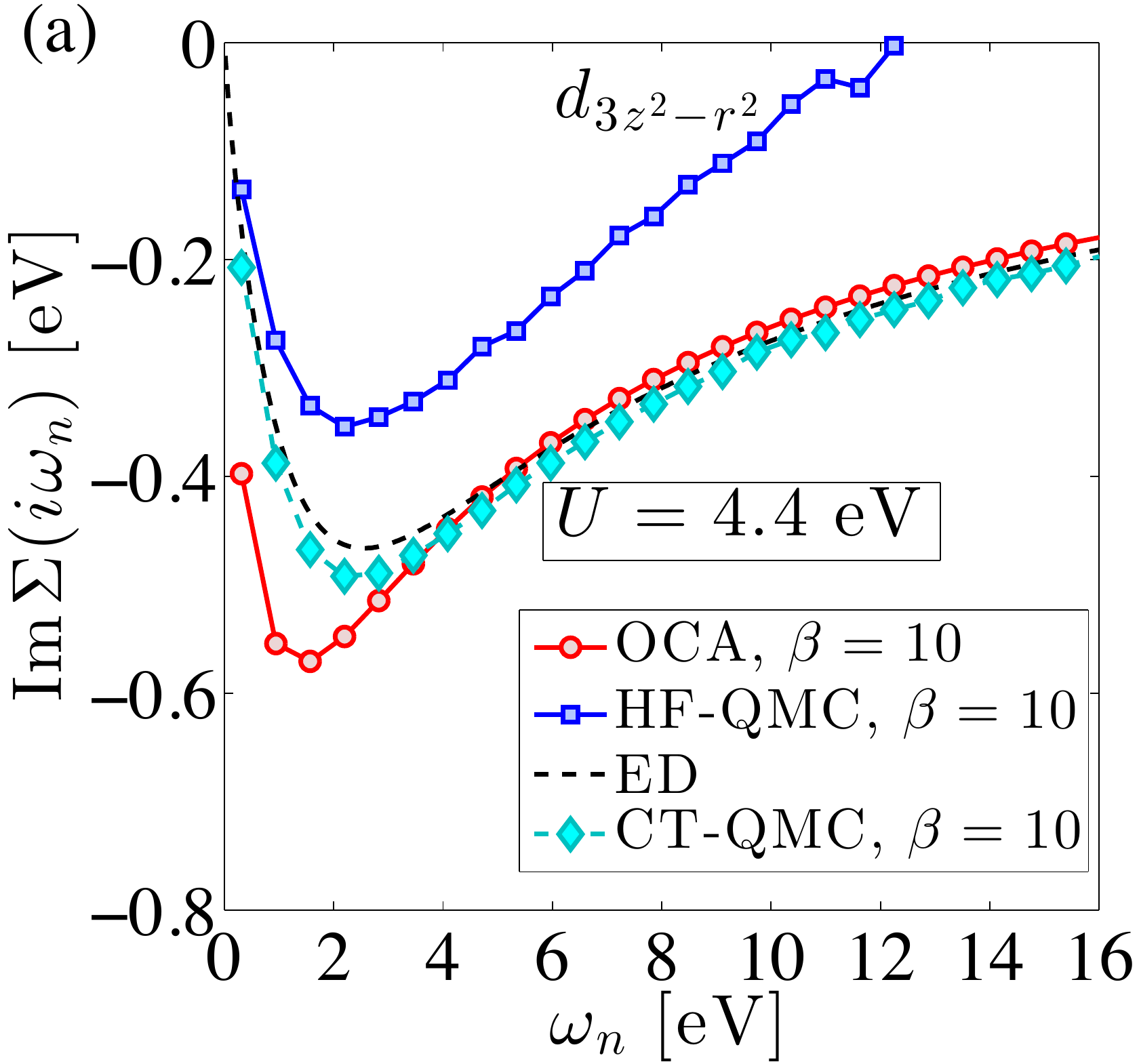}
\includegraphics[width=0.49\linewidth]{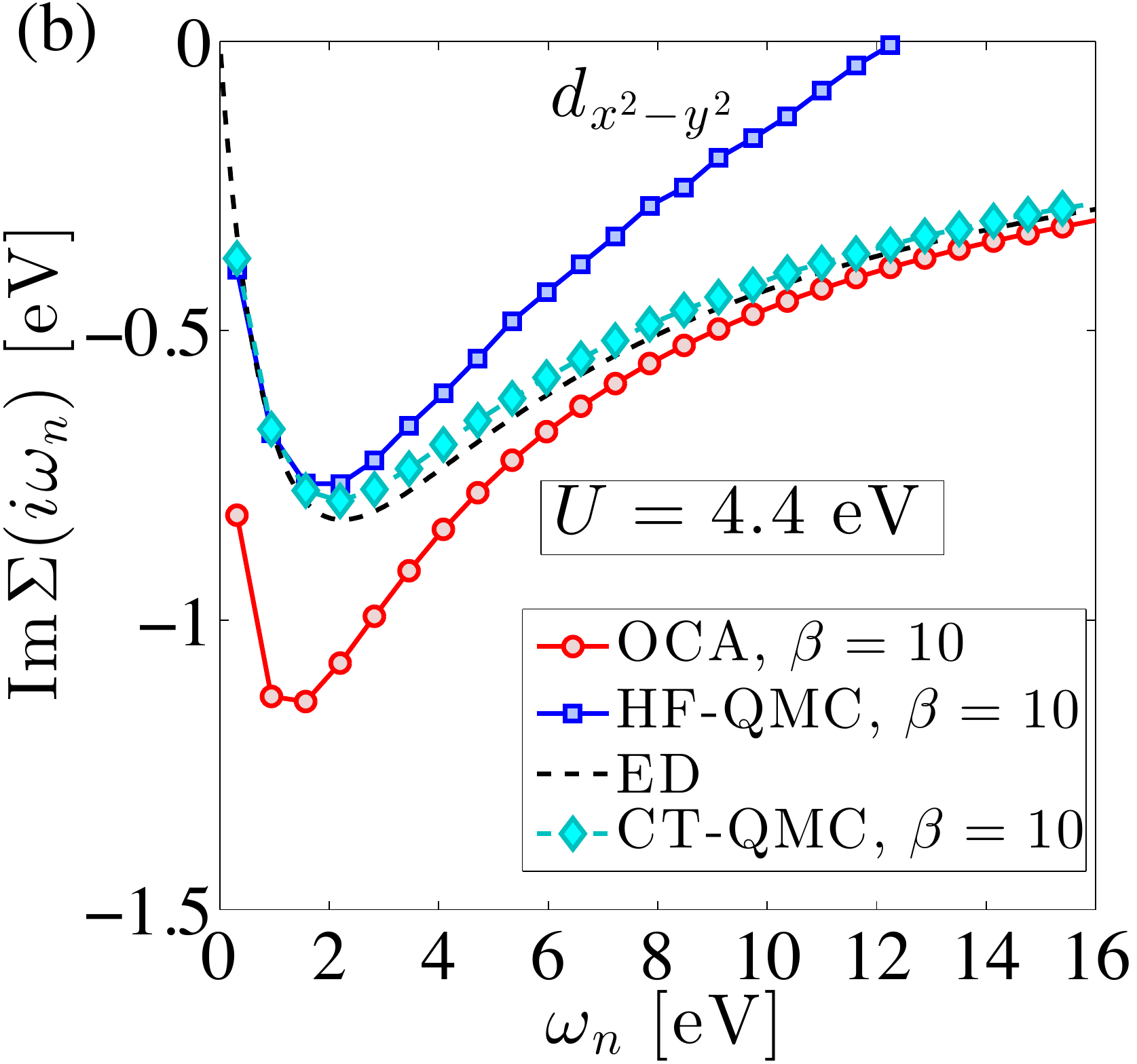}
\includegraphics[width=0.49\linewidth]{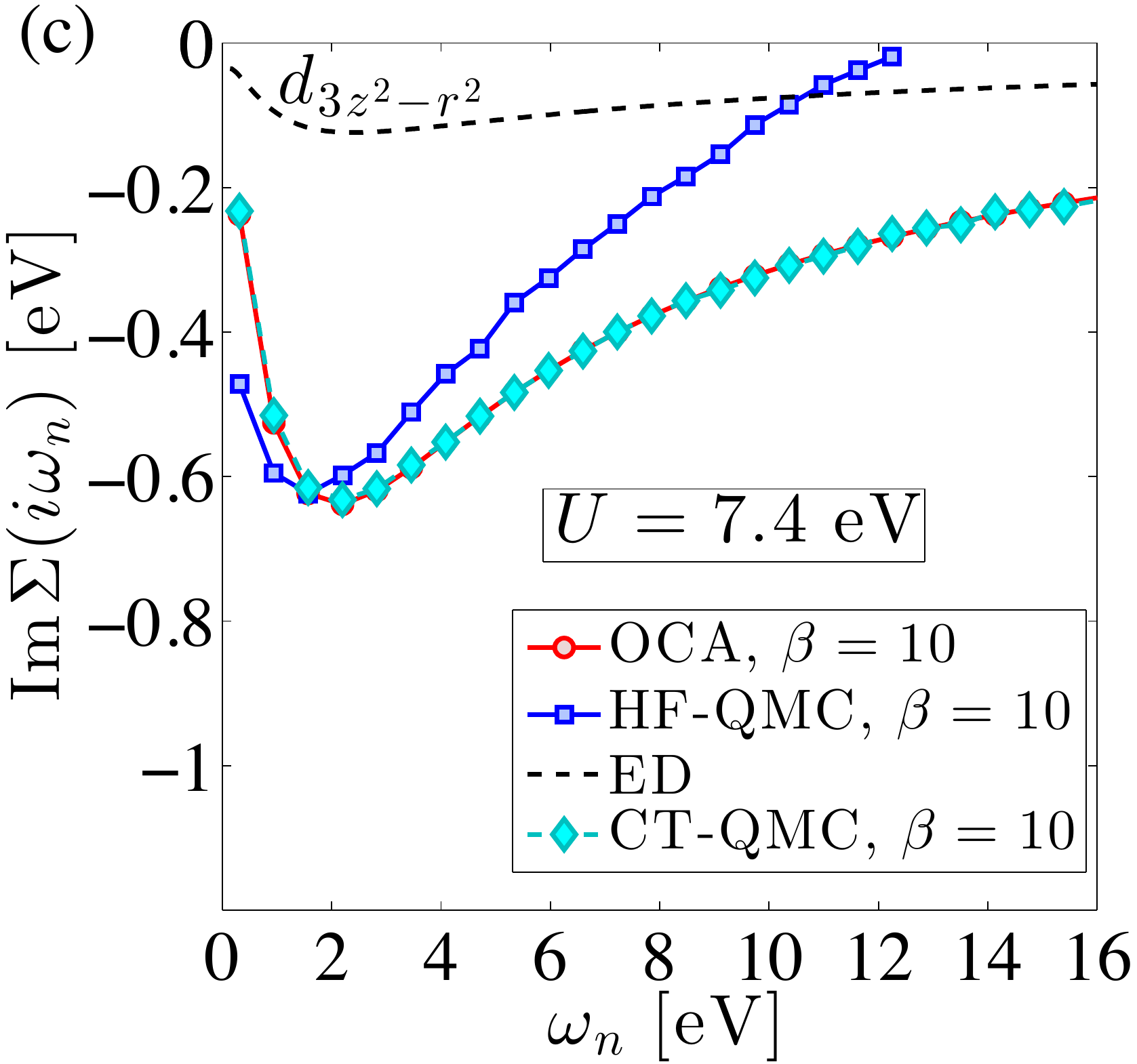}
\includegraphics[width=0.49\linewidth]{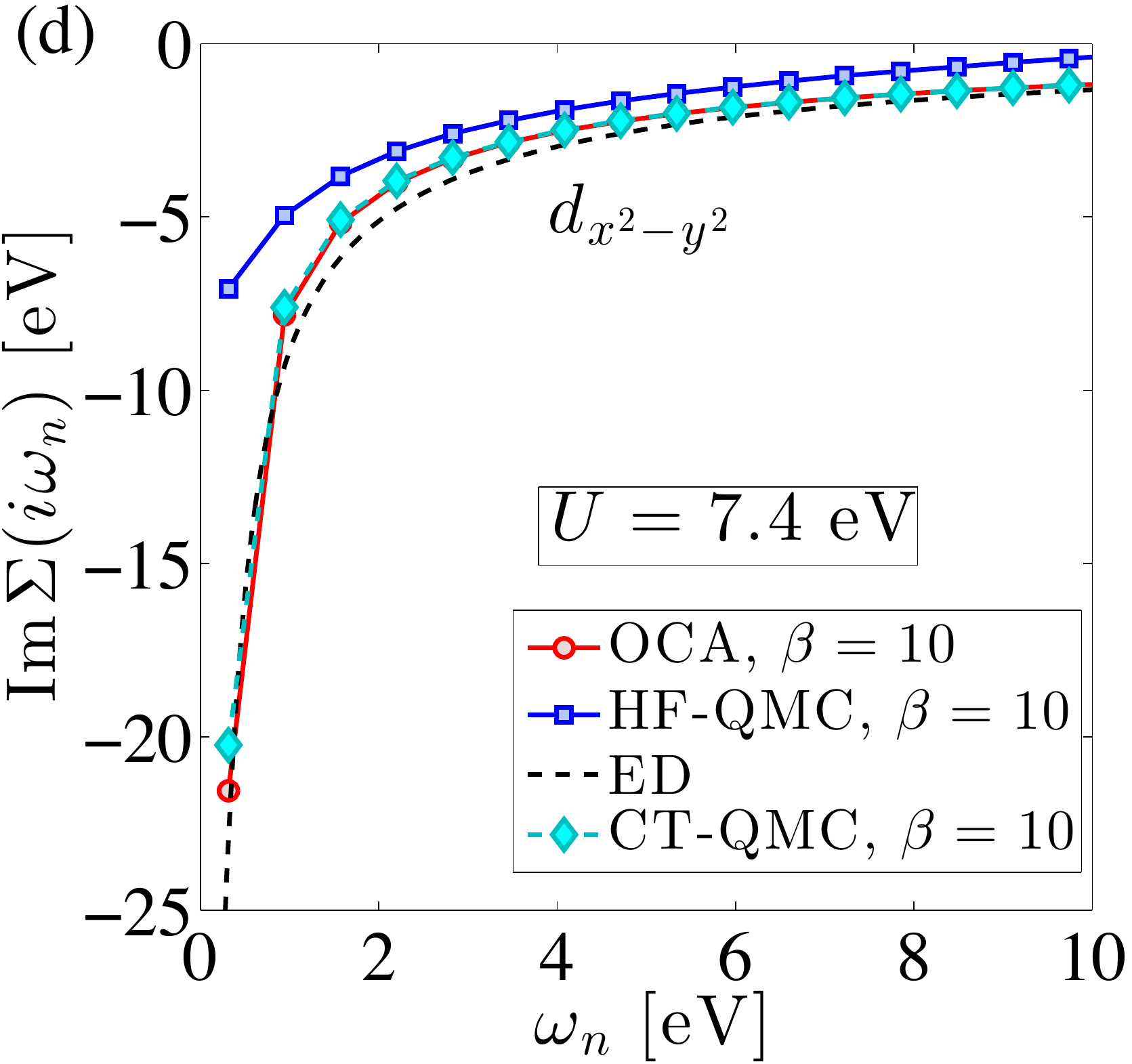}
\caption{Comparison of the imaginary part of the self-energy for the $d_{3z^2-r^2}$ orbital [(a) and (c)] and the $d_{x^2-y^2}$ orbital [(b) and (d)] as function of Matsubara frequency obtained with different impurity solvers: OCA, ED and CT-QMC. Also shown are data from Ref.~\onlinecite{Hansmann:2010} obtained within HF-QMC. (a) and (b) are in the metallic state with $U=4.4$~eV and (c) and (d) are in the insulating state with $U=7.4$ eV. The inverse temperature for QMC and OCA is set at $\beta=10$~eV$^{-1}$ while ED results are obtained using 5 bath sites per orbital and a fictive temperature $T_0=0.0005t$.}
\label{fig:ImSigma}
\end{figure}

We start with the smaller value of the interaction, $U=4.4$ eV, where a metallic state is observed. Figure~\ref{fig:ImSigma} shows the imaginary part of the self-energy for (a) the $d_{3z^2-r^2}$ and (b) the $d_{x^2-y^2}$ orbital obtained with the OCA and ED solvers along with the QMC results. The OCA self-energy agrees well with the CT-QMC in the high frequency regime but overestimates the magnitude of the self-energy for small frequencies. This overestimation of correlation effects reflects the fact that the OCA tends to favor the insulating state. The ED results agree with the CT-QMC over the full range of Matsubara frequencies. This suggests that the low-energy properties of the metallic state are reliably obtained within ED. On the other hand, the HF-QMC results of Ref.~\onlinecite{Hansmann:2010} show a systematic deviation from our results at larger frequencies. The HF-QMC is numerically exact, only if the extrapolation of the imaginary-time slice $\Delta\tau$ to zero has been done. Otherwise, it provides an approximate solution. More specifically, the data of Ref.~\onlinecite{Hansmann:2010} were obtained from a HF-QMC code which uses the ``Ulmke smoothing" to adjust the high-frequency tails. This introduces an additional systematic error.\footnote{G.~Sangiovanni, private communication (2014).}

\begin{figure}
\includegraphics[width=0.485\linewidth]{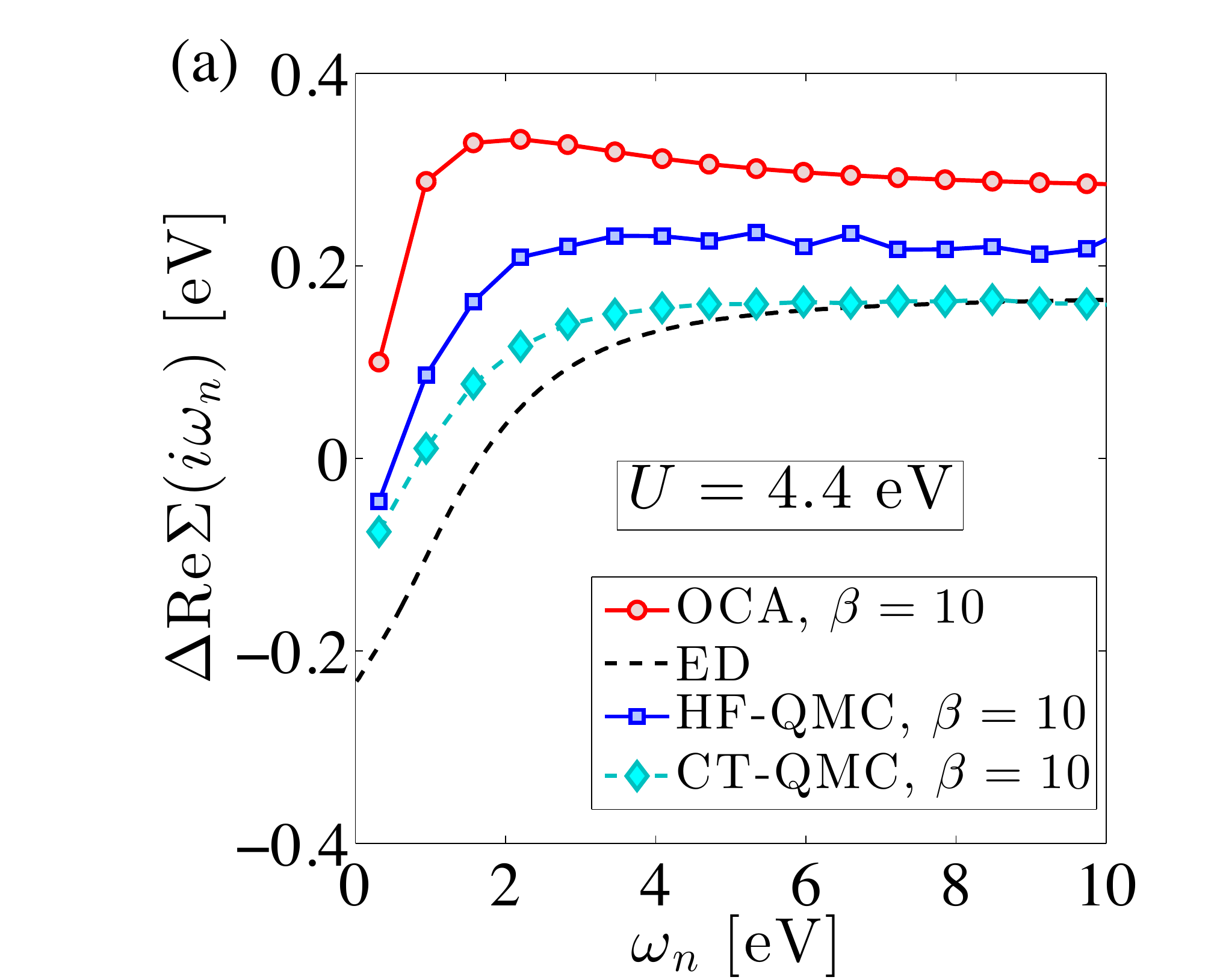}
\includegraphics[width=0.495\linewidth]{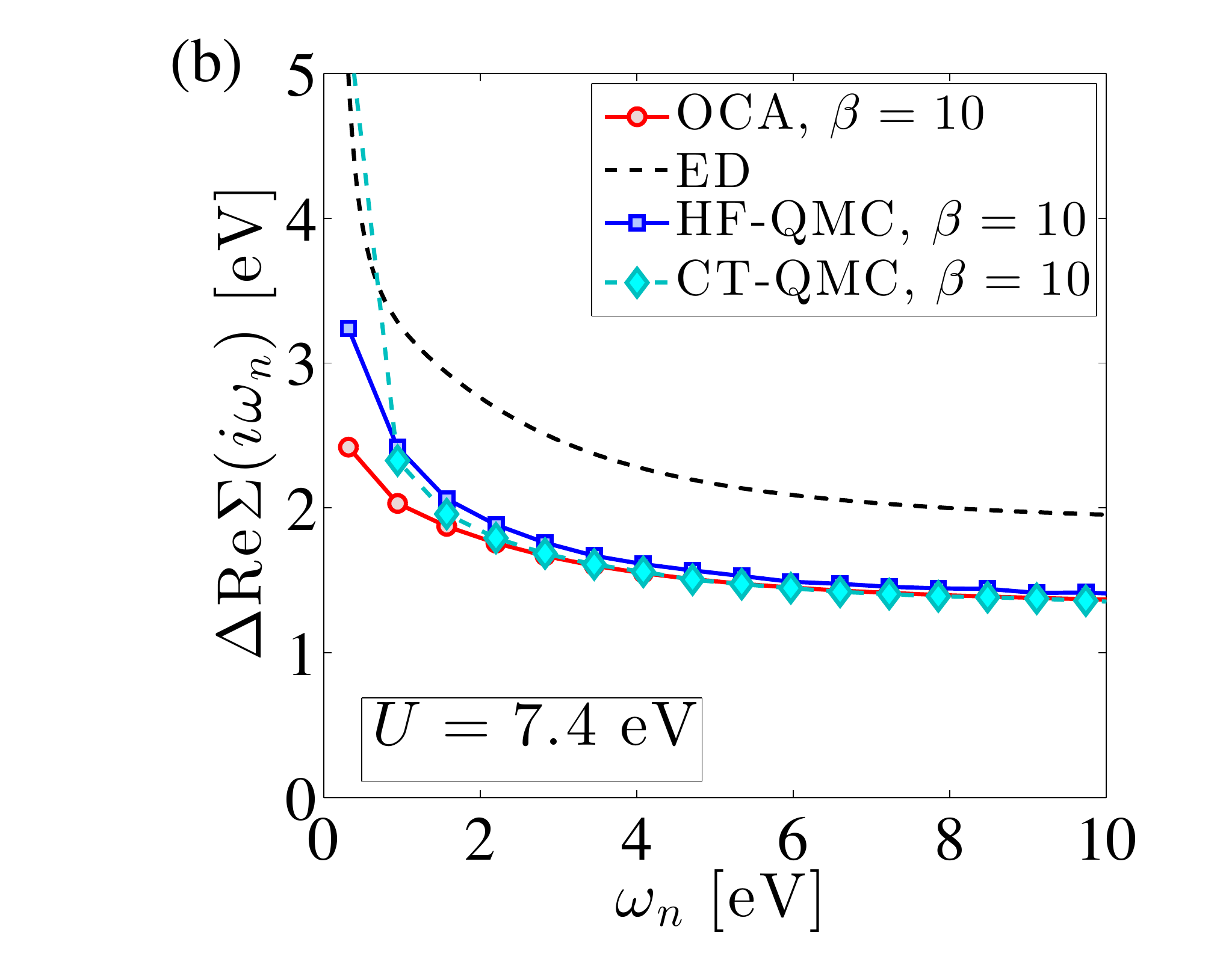}
\caption{Orbital difference of the real part of the self-energy $\Delta{\rm Re}\,\Sigma(i\omega_n)={\rm Re}\,\Sigma_{3z^2-r^2}(i\omega_n)-{\rm Re}\,\Sigma_{x^2-y^2}(i\omega_n)$ for the same parameters as in Fig.~\ref{fig:ImSigma}.}
\label{fig:ReSigma}
\end{figure}

The difference $\Delta{\rm Re}\,\Sigma(\omega_n)={\rm Re}\,\Sigma_{3z^2-r^2}(\omega_n)-{\rm Re}\,\Sigma_{x^2-y^2}(\omega_n)$ of the real part of the self-energy between the two orbitals in the metallic state is shown in Fig.~\ref{fig:ReSigma}(a). The general trend among the different solvers is consistent but the OCA shows a systematic shift in the high-frequency limit. This difference is also manifest in the violation of sum rules known to occur in the OCA\cite{Ruegg:2013b} and will be discussed in Sec.~\ref{sec:sum-rule}.

We now turn to the insulating case at $U=7.4$ eV. Figure~\ref{fig:ImSigma}(c) and (d) show the imaginary part of the Matsubara self-energy for the two $e_g$ orbitals as obtained from the different impurity solvers. This comparison confirms the expectation that the OCA performs better in the insulating phase: within the accuracy of Fig.~\ref{fig:ImSigma}, the OCA self-energy is indistinguishable form the CT-QMC. A similar conclusion can be drawn from the real part of the self-energy shown in Fig.~\ref{fig:ReSigma}(b). The ED shows a behavior similar to CT-QMC/OCA for ${\rm Im}\,\Sigma(i\omega_n)$ of the strongly correlated $d_{x^2-y^2}$ orbital [Fig.~\ref{fig:ImSigma}(d)] while it predicts a smaller value of the imaginary part of the self-energy for the $d_{3z^2-r^2}$ orbital [Fig.~\ref{fig:ImSigma}(c)]. We attribute this difference to a temperature effect: as $T$ approaches zero, the $d_{3z^2-r^2}$ orbital gets depleted almost entirely which leads to a small self-energy for the $d_{3z^2-r^2}$ orbital. This effect is also visible in the real part of the self-energy, as shown in Fig.~\ref{fig:ReSigma}(b). We note that we found a consistent trend by performing CT-QMC calculations at lower temperatures.

\subsection{Sum rules within OCA}
\label{sec:sum-rule}
\begin{figure}
\includegraphics[width=0.49\linewidth]{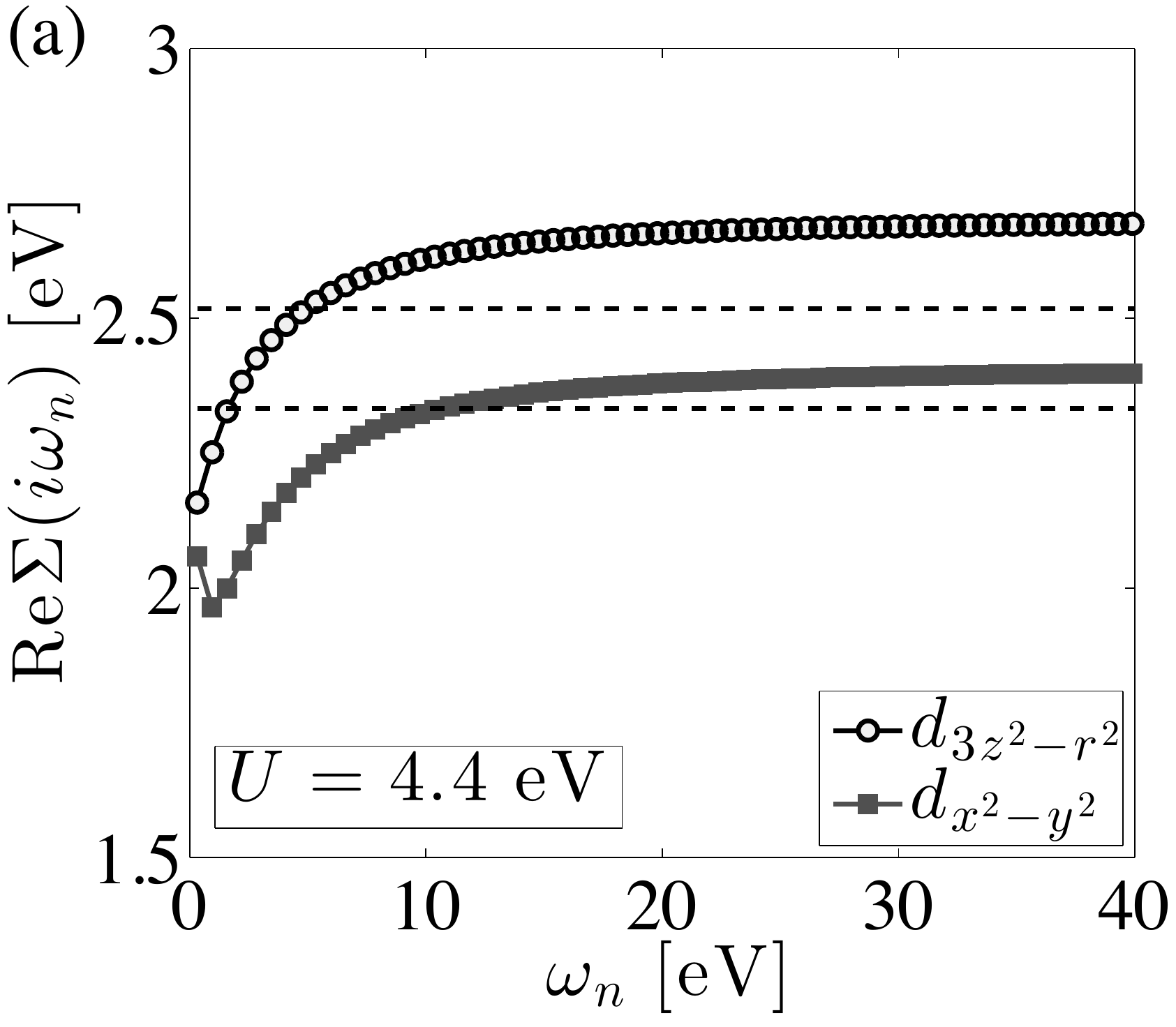}
\includegraphics[width=0.49\linewidth]{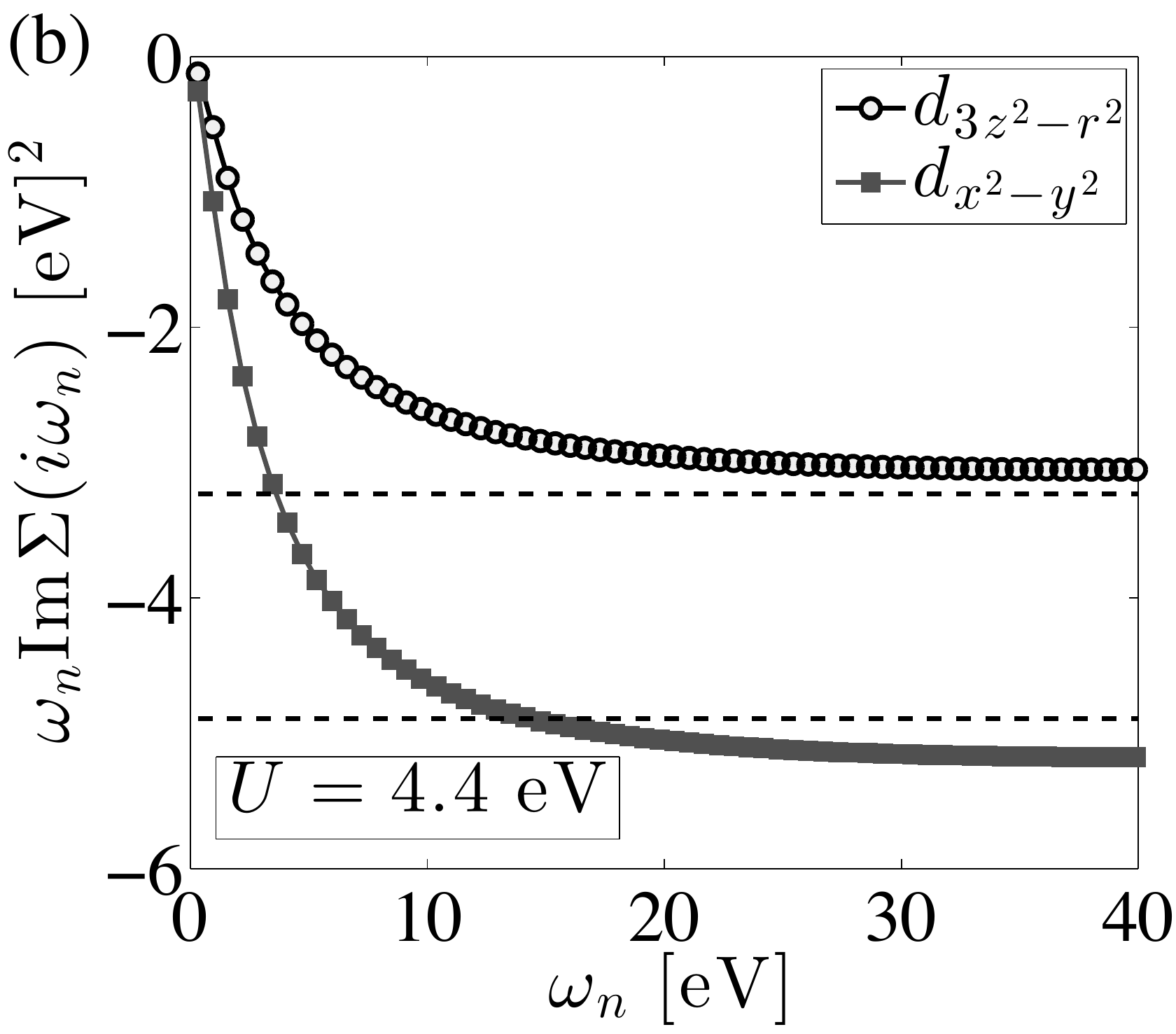}
\includegraphics[width=0.49\linewidth]{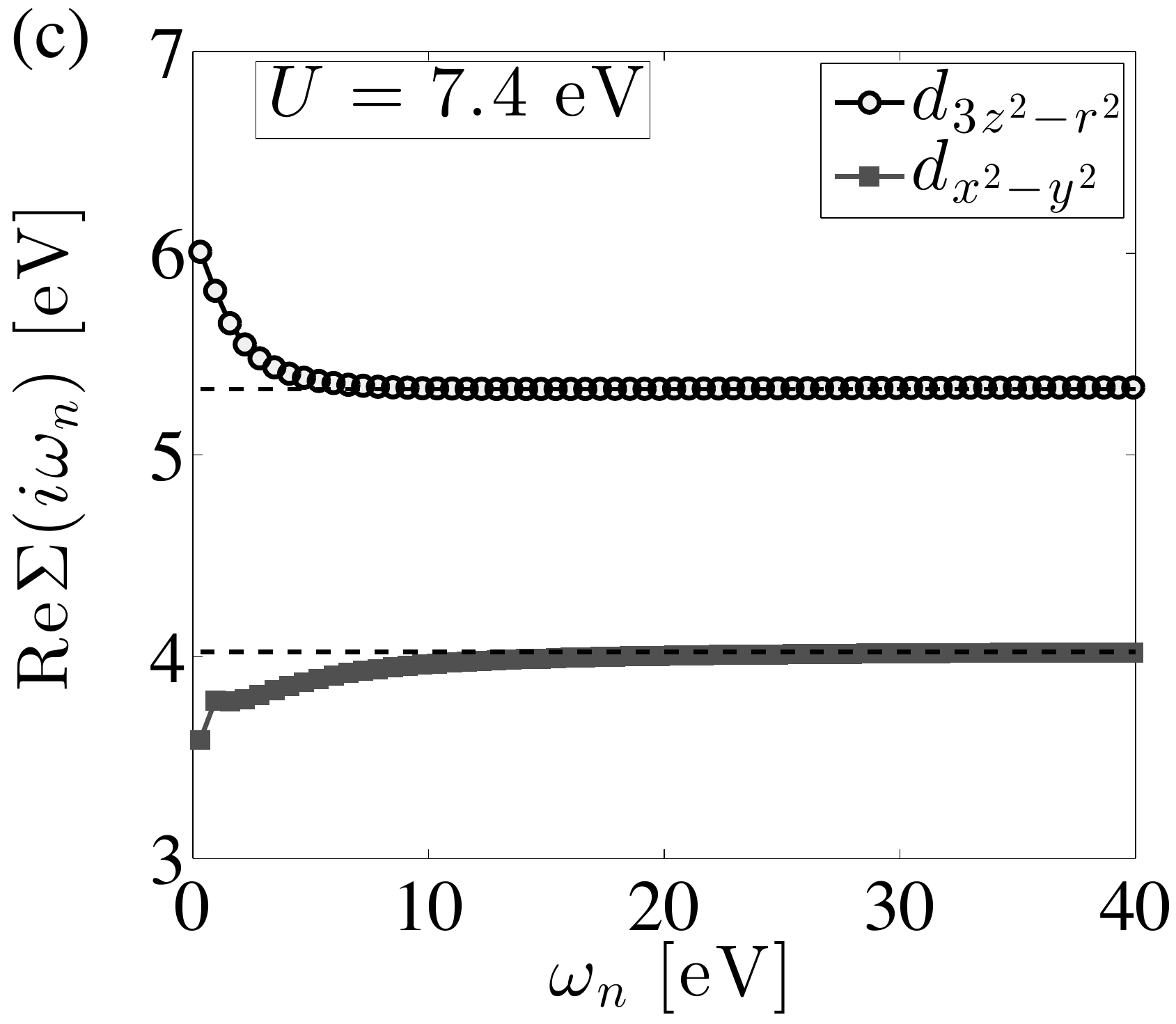}
\includegraphics[width=0.49\linewidth]{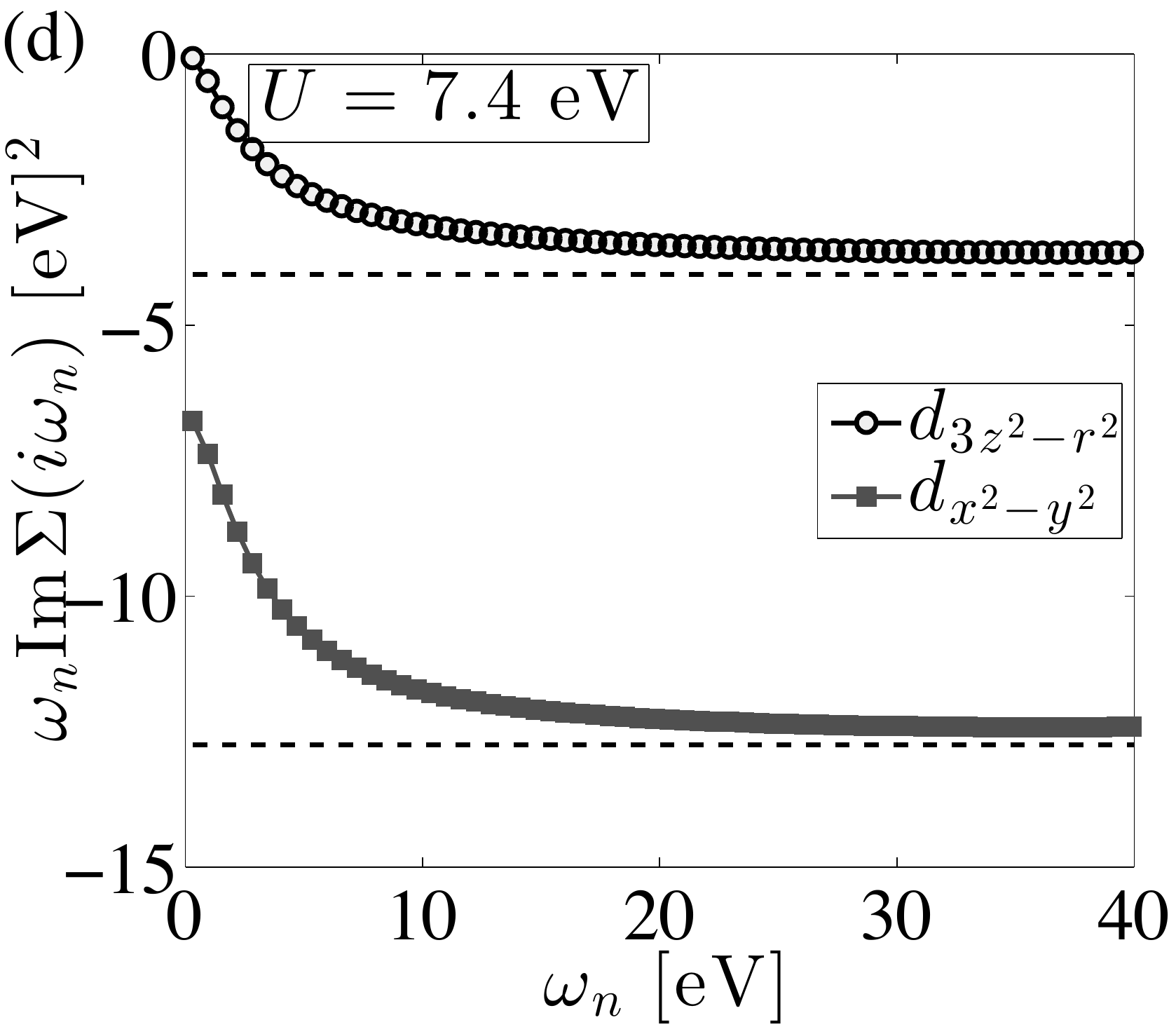}
\caption{The frequency dependence of ${\rm Re}\,\Sigma(i\omega_n)$ [(a) and (c)] and $\omega_n{\rm Im}\Sigma(i\omega_n)$ [(b) and (d)] for the two orbitals obtained within DMFT(OCA) for $U=4.4$ eV [(a) and (b)] and $U=7.4$ eV [(c) and (d)]. The dashed lines indicate the value for the high-frequency limit if the sum rules were satisfied, see main text. The deviation serves as an internal consistency check to address the quality of the OCA.}
\label{fig:sumrule}
\end{figure}

As a further test for the accuracy of the DMFT(OCA), we briefly discuss the sum rule violations encountered for the self-energy.
In Ref.~\onlinecite{Ruegg:2013b}, it was argued that the degree to which these sum rules are violated in the OCA provides an internal self-consistency test and can be used to estimate the quality of the approximation. Specifically, let us consider the high-frequency expansion of the electronic self-energy:
\begin{equation}
\Sigma(i\omega_n)=\Sigma_0+\frac{\Sigma_1}{i\omega_n}+\frac{\Sigma_2}{(i\omega_n)^2}+\dots.
\end{equation}
Exact identities relate the coefficients in the above expansion to thermodynamic expectation values of certain commutators of the Hamiltonian, see e.g.~Ref.~\onlinecite{Potthoff:1997}. However, these sum rules are in general violated within the OCA and it was argued\cite{Ruegg:2013b} that the degree of the sum rule violation yields an estimate of the overall accuracy of the OCA. Figure~\ref{fig:sumrule} compares the high frequency limit of (a) ${\rm Re}\,\Sigma(i\omega_n)$ and (b) $\omega_n{\rm Im}\, \Sigma(i\omega_n)$ to the values expected from the sum rules (dashed lines) in the metallic phase at $U=4.4$ eV. Similarly, (c) and (d) show the same quantities in the insulating phase at $U=7.4$ eV. As expected, the violations of the sum rules are smaller in the insulating phase, suggesting a higher accuracy of the OCA for larger interactions which is in agreement with the direct comparison to the CT-QMC results.
\section{Results}
\label{sec:results}
In the next two sections, we discuss the dependence of the electronic structure on the number of layers $L$ in the thin film geometry as well as the dependence on the carrier concentration. In Sec.~\ref{sec:layer}, we fix the carrier concentration at quarter filling, $n=1$, and study the layer-resolved orbital polarization. We also obtain the metal-insulator phase diagram for different thicknesses. In Sec.~\ref{sec:carrier}, we concentrate on the single-layer system $L=1$ and investigate the influence of varying the carrier concentration. Throughout this section, we consider the simplest version of the two-orbital model Eq.~\eqref{eq:H} with $t'=\Delta=0$ and measure energies in units of the (single) nearest-neighbor hopping amplitude $t$.
\subsection{Dependence on layer thickness for $n=1$}
\label{sec:layer}
\begin{figure}
\includegraphics[width=1\linewidth]{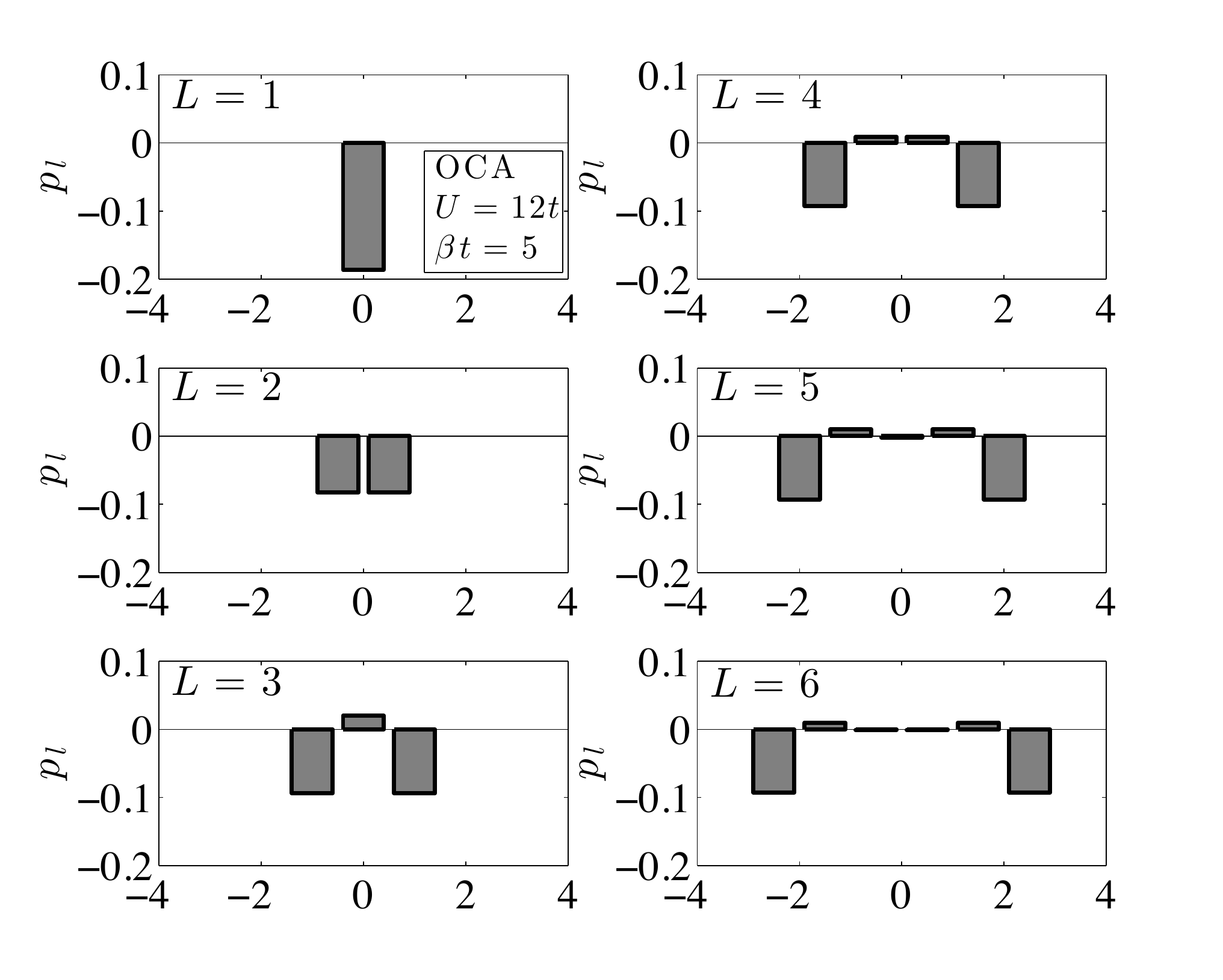}
\caption{Layer-dependent polarization for thin films with $L=1$ to $L=6$ atomic layers at quarter filling obtained within DMFT(OCA). The interaction parameters have been fixed at $U=12t$ and $J=t$ and the inverse temperature at $\beta t=5$.}
\label{fig:polarization_OCA}
\end{figure}
\begin{figure}
\includegraphics[width=1\linewidth]{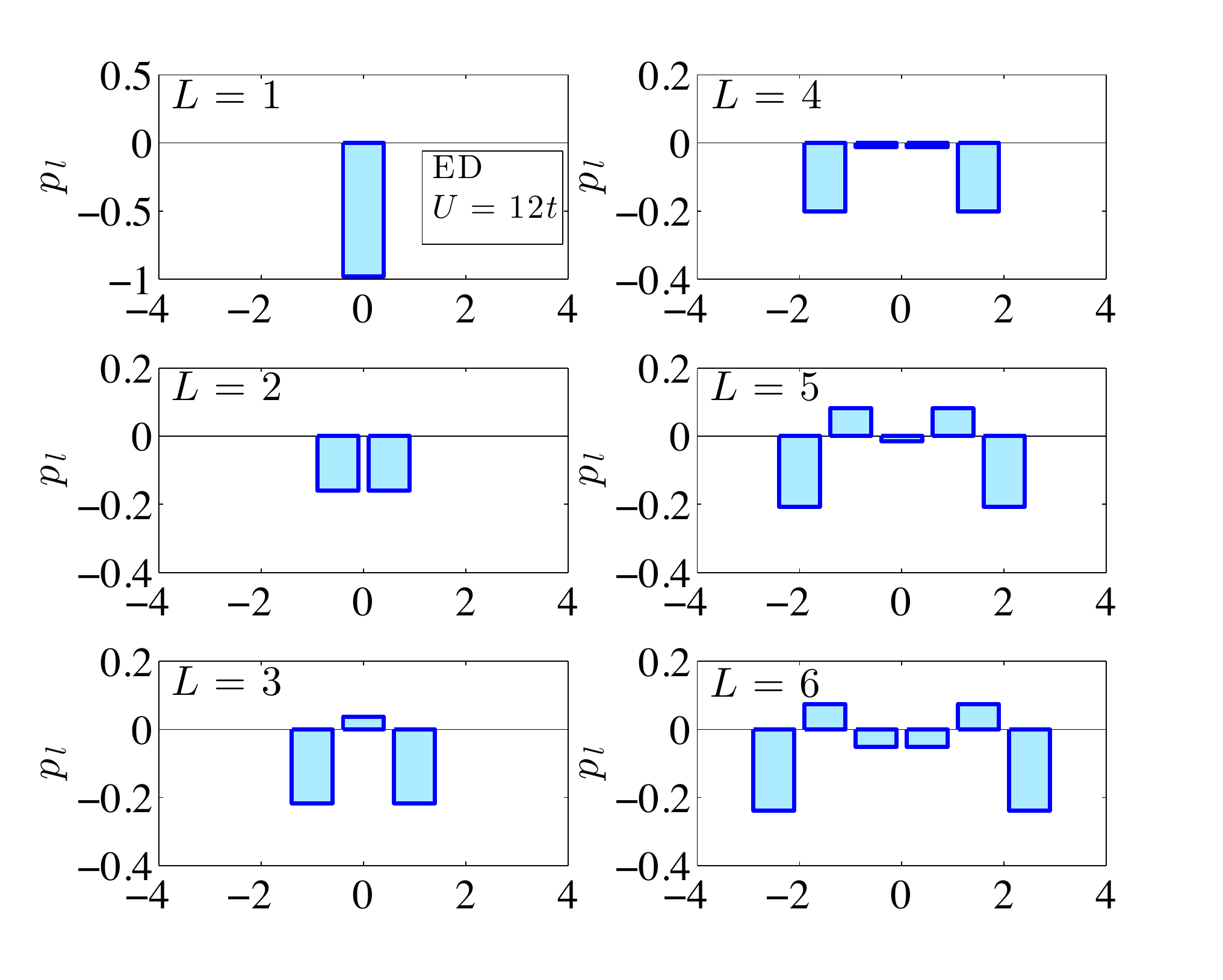}
\caption{Layer-dependent polarization for thin films with $L=1$ to $L=6$ atomic layers at quarter filling obtained within DMFT(ED) using 5 bath sites per orbital and a fictive temperature $T_0=0.0005t$. The interaction parameters have been fixed at $U=12t$ and $J=t$.}
\label{fig:polarization_ED}
\end{figure}

We have solved the layer-DMFT self-consistency Eq.~\eqref{eq:DMFT} for systems up to $L=6$ layers using the OCA and ED solvers. Figures~\ref{fig:polarization_OCA} and \ref{fig:polarization_ED} show the layer-resolved orbital polarization for thin films of various thicknesses deep in the Mott insulating phase ($U=12t$), where we expect our approximate solvers to work fine. The local orbital polarization
\begin{equation}
p_l=n_{3z^2-r^2,l}-n_{x^2-y^2,l},
\end{equation}
measures the difference of the orbital occupation in layer $l$. It has been demonstrated that the spatially resolved orbital polarization can be obtained experimentally using soft-X-ray reflectometry.\cite{Benckiser:2011,Wu:2013} In Sec.~\ref{sec:conclusions}, we discuss our theoretical results in view of these experiments. 

For all the thin films we have studied, we find a sizable (negative) orbital polarization for the interface layers, indicating the preference to occupy the $d_{x^2-y^2}$ orbital. As mentioned previously, the origin of the orbital polarization lies in the reduced symmetry of the (001) films. We emphasize that our model does not include an explicit crystal-field splitting which would affect the orbital occupation in the atomic limit. Instead, in the thin film geometry, the kinetic energy of the $d_{3z^2-r^2}$ electrons is quenched as compared to the $d_{x^2-y^2}$ electrons. To optimize the kinetic energy, electrons preferably occupy the $d_{x^2-y^2}$ orbital, thus building up an orbital polarization. In the DMFT calculation, the reduced symmetry manifests itself as an orbitally asymmetric hybridization function. For increasing thickness $L$, the orbital polarization in the center of the structure approaches a vanishingly small value. This is in agreement with our bulk calculations where we find zero orbital polarization in the Mott insulator at quarter filling. However, because of the local symmetry breaking, $p_l\neq 0$ in general. 

The comparison between the OCA data at $\beta t=5$ (Fig.~\ref{fig:polarization_OCA}) and the ED data (Fig.~\ref{fig:polarization_ED}) reveals an overall consistent behavior. It also shows a clear temperature dependence of the polarization: as expected, the magnitude increases as $T\rightarrow 0$. Finally, we remark that ED data were obtained including $N_h=5$ bath sites per orbital. We have also studied $N_h=3$ and 4 and the behavior of the polarization as function of $N_h$ suggests that convergence with respect to $N_h$ is relatively slow for the considered systems. More details are provided in the Appendix.

\begin{figure}
\includegraphics[width=0.8\linewidth]{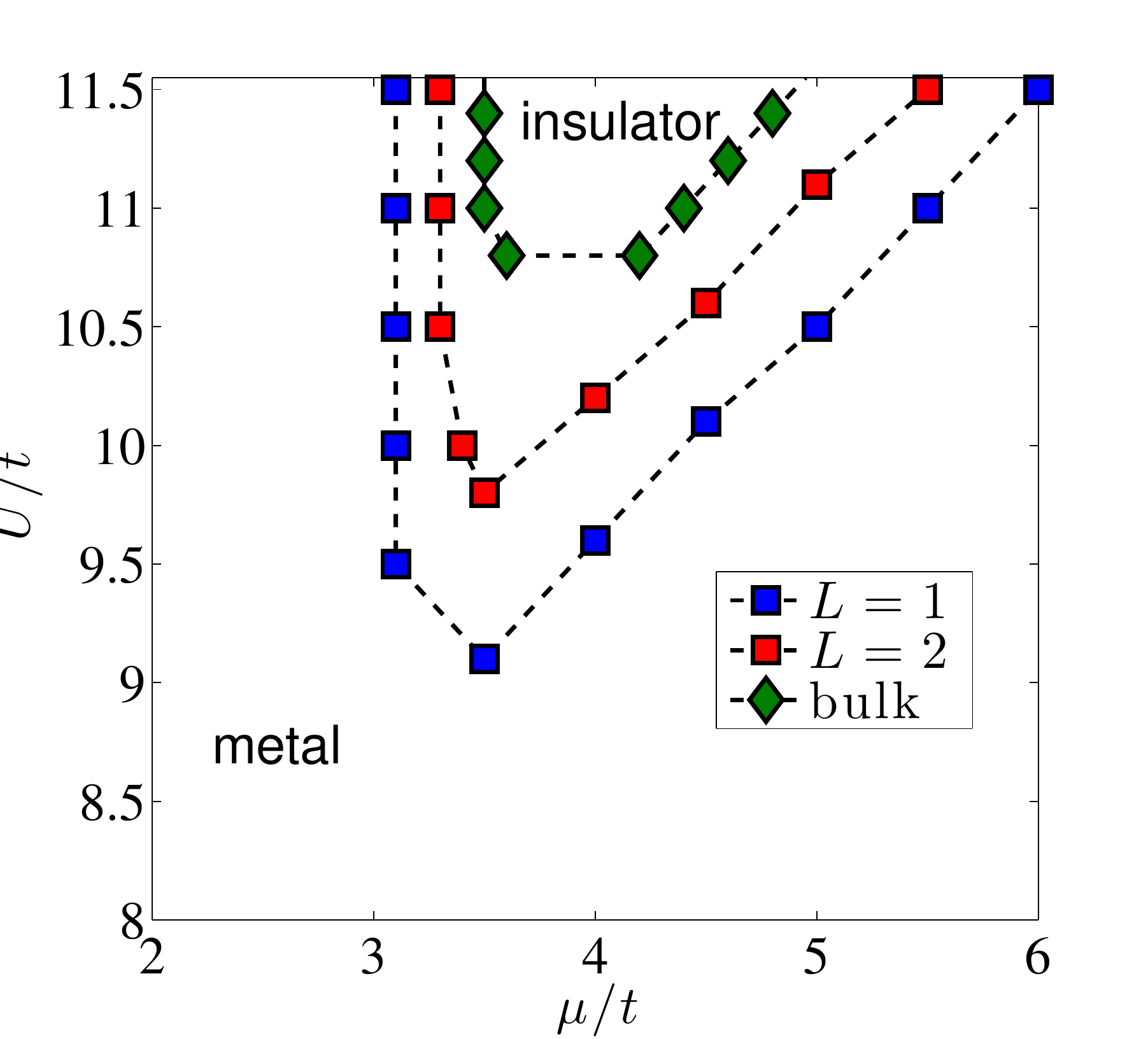}
\caption{Metal-insulator phase diagram in the $\mu$-$U$ plane near quarter filling for $L=1$, $L=2$ and bulk for fixed $J=t$ using DMFT(OCA). The phase boundaries were determined from the width of the charge plateaus in the $n(\mu)$ curves at $\beta t=10$.}
\label{fig:MIT}
\end{figure}
The dimensional reduction inherent in the few-layer system also affects the total kinetic energy of the electrons in the thin film, leading to an enhancement of correlation effects in the very thin limit.\cite{Potthoff:1999} In particular, the location of the metal-insulator transition/crossover found in bulk is modified in the few-layer systems. As shown in Fig.~\ref{fig:MIT}, the insulating phase in the $L=1$ system is considerably larger than in bulk. For $L=2$, the phase boundary is shifted towards the bulk and increasing the number of layers even further, we expect that the bulk phase boundary is rapidly approached. 

The increased stability of the insulating phase due to quantum confinement is in qualitative agreement with recent experimental results on (LaNiO$_3$)$_n$/(LaAlO$_3$)$_N$ superlattices where an insulating phase has been found in the thin limit with $n\leq3$.\cite{Boris:2011,Liu:2011} Furthermore, anti-ferromagnetic order was identified below $T_N\approx50^\circ K$ for the insulating superlattices.\cite{Boris:2011} Our theoretical phase diagram was obtained assuming a paramagnetic Mott insulator, which is a reasonable assumption at the elevated temperatures used in Fig.~\ref{fig:MIT}. However, we expect that the {\em ground-states} of the insulating systems will also develop magnetic order within the DMFT framework. We also remark that the precise nature of the insulating state observed in experiment is still a matter of active research, see e.g.~Refs.~\onlinecite{Freeland:2011,Park:2012}, and goes beyond the scope of the present work.

\subsection{Dependence on carrier density for $L=1$}
\label{sec:carrier}
\subsubsection{Orbital polarization}
\label{sec:OP}

%
\begin{figure}
\includegraphics[width=0.7\linewidth]{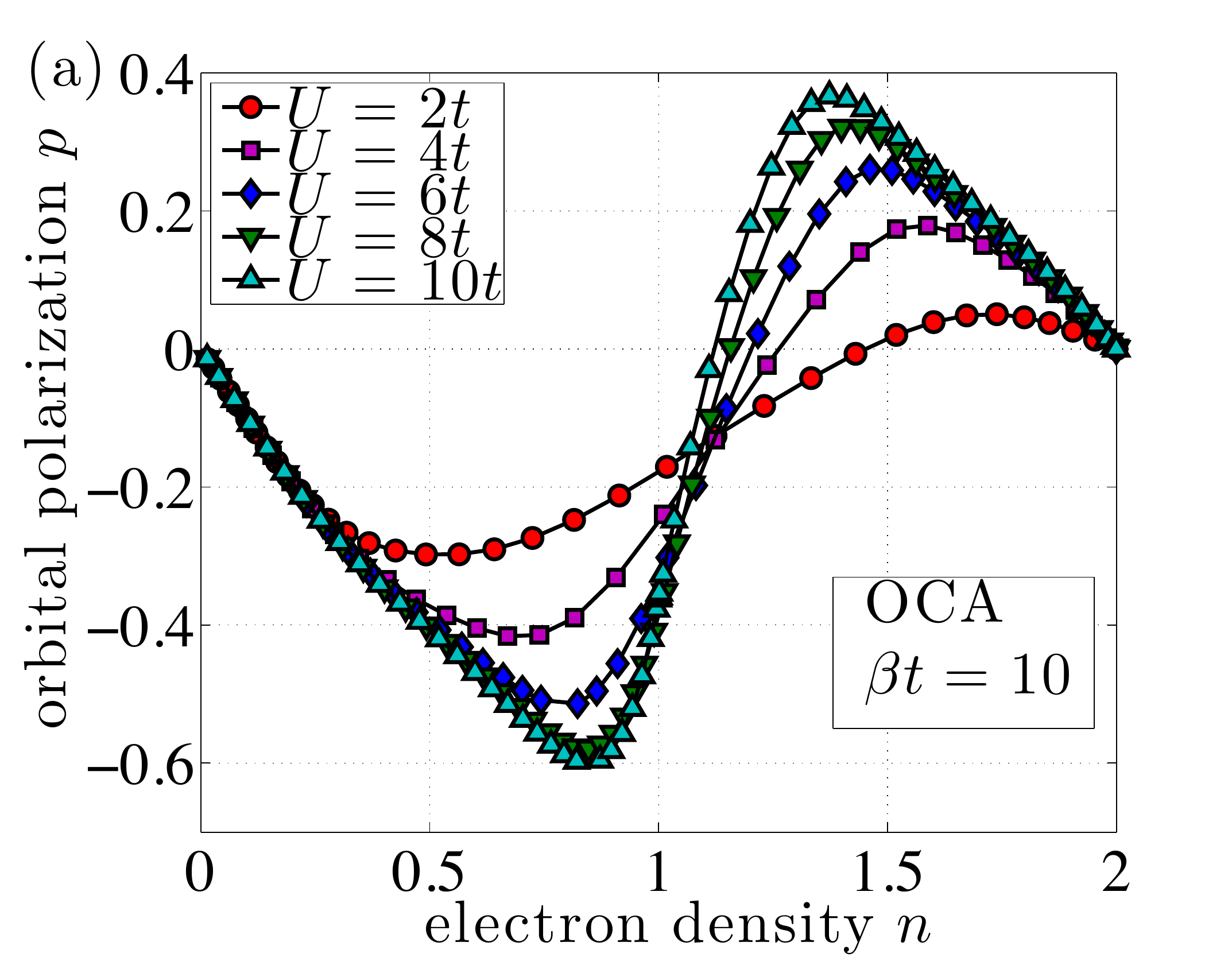}
\includegraphics[width=0.7\linewidth]{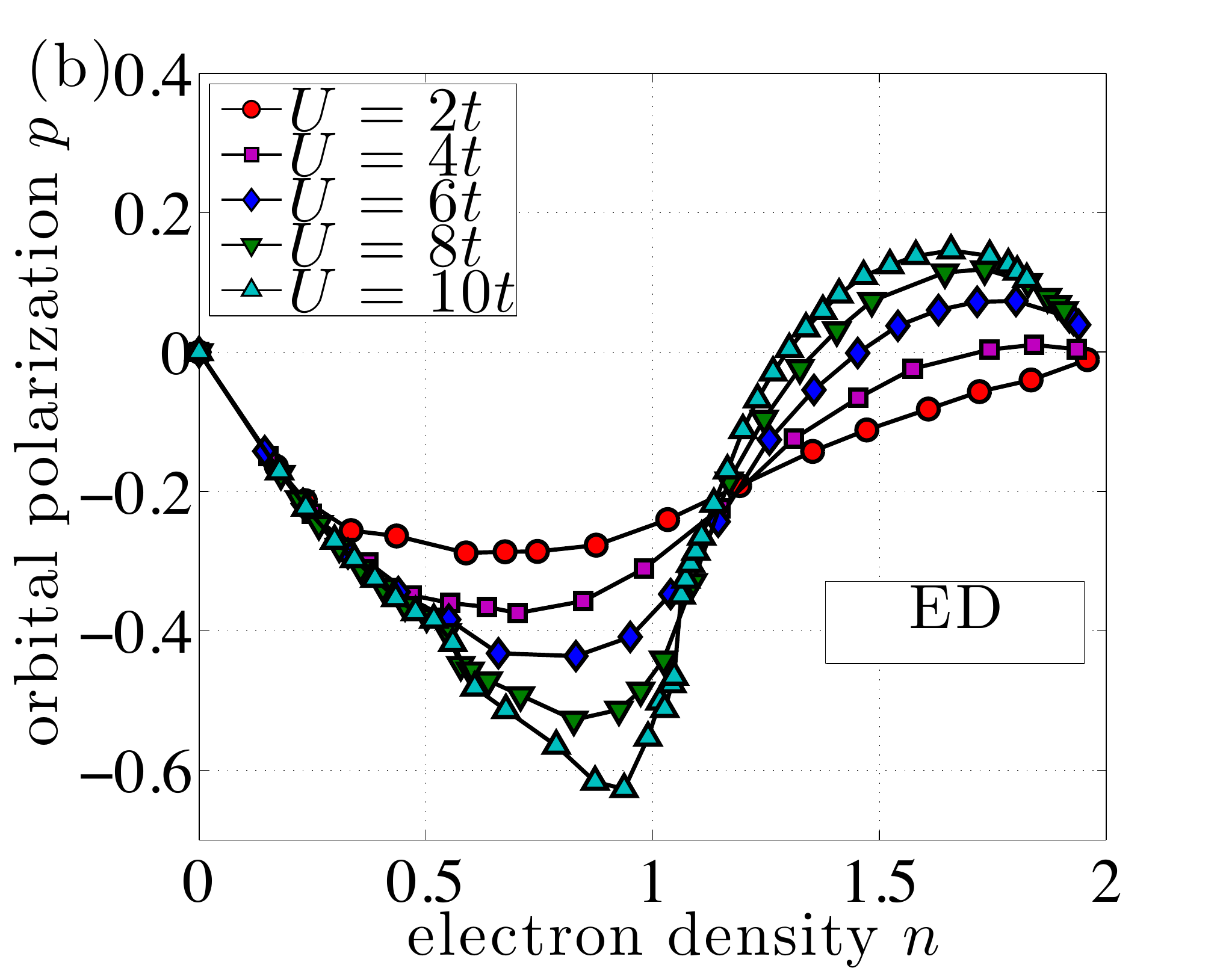}
\includegraphics[width=0.7\linewidth]{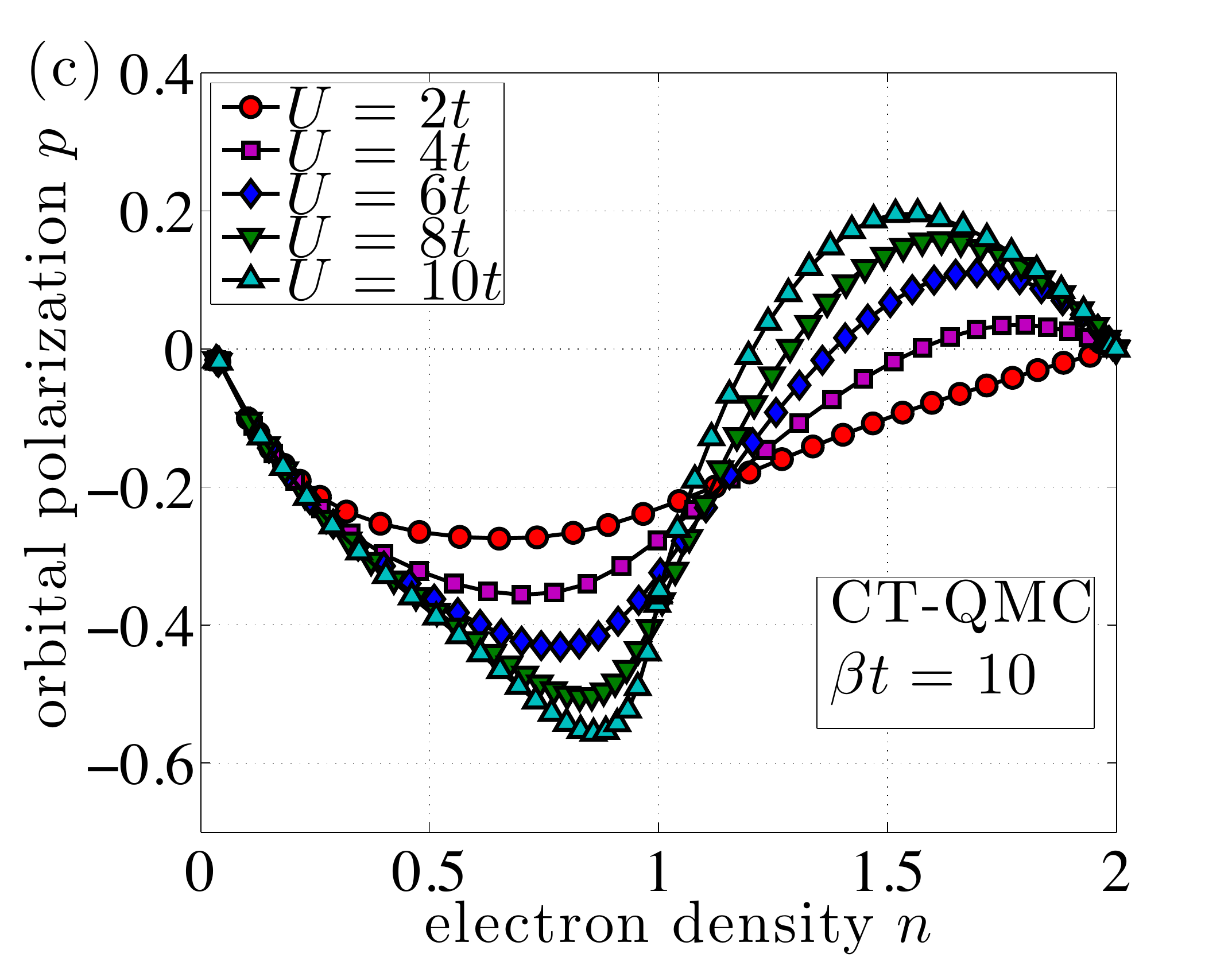}
\caption{The dependence of the orbital polarization $p$ on the electron density $n$ in the monolayer model ($L=1$) for $\beta t=10$ within (a) DMFT(OCA) and (c) DMFT(CT-QMC). In (b), the orbital polarization obtained within DMFT(ED) using 5 bath sites per orbital and a fictive temperature $T_0=0.0005t$ is shown. Different curves represent different interaction strengths $U=2,4,6,8,10t$ at fixed Hund's coupling $J=t$.}
\label{fig:polarization_n}
\end{figure}

We next investigate how physical quantities depend on the carrier concentration $n$ for the single layer model with $L=1$, where the effect of the local symmetry-breaking is strongest. We first focus on the orbital polarization $p$. Figures~\ref{fig:polarization_n}(a)-(c) show the dependence of $p$ on $n$ for various interaction strengths as obtained within OCA, ED and CT-QMC, respectively. The results obtained using the different impurity solvers qualitatively agree with each other. Namely, $p$ depends quite strongly on $n$: in particular, while $p$ below and around quarter filling ($n=1$) is negative, it assumes positive values in the vicinity of $n=1.5$ for large interactions. Such positive values of $p$ result from correlation effects and are absent for vanishing interactions. Turning to a quantitative comparison, one identifies differences between OCA and CT-QMC. For example, as compared to the numerically exact CT-QMC values, OCA overestimates the magnitude of the polarization in the vicinity of $n=1.5$ by roughly a factor two. The discrepancy is similarly pronounced for both small and large interactions and it reveals a shortcoming of the OCA: even if interactions are large, the accuracy of the OCA is reduced as soon as the system is tuned away from the insulator. On the contrary, ED gives results which are consistent with the exact results at $\beta t=10$.

In the strongly correlated limit, we can understand the sign of the polarization from a simple physical picture. Near the empty band limit $n=0$, carriers are electron-like and predominantly occupy the $d_{x^2-y^2}$ orbital in order to optimize their kinetic energy. This results in $p<0$. On the other hand, approaching $n=2$, the mobile carriers are holes, which are doped into the Mott insulator. In order to optimize their kinetic energy, they also occupy the $d_{x^2-y^2}$ orbital. This means that more electrons reside in the $d_{3z^2-r^2}$ orbital which results in a positive orbital polarization. From these considerations, one expects that the orbital polarization for densities slightly above $n=2$ is again negative in the strongly interacting limit. Indeed, we have numerically confirmed this expectation for $U=8t$ and $U=10t$. In fact, because the considered model with only nearest-neighbor hopping is particle-hole symmetric, $p(n)$ is odd around half-filling, i.e.~$p(2+x)=-p(2-x)$ where $-2<x<2$ measures the density from half filling. This relation forces $p(n=2)=0$ and allows one to obtain the polarization for $n$ between 2 and 4 for arbitrary interactions.

The dependence of the orbital polarization on the carrier density indicates an interesting renormalization of the (implicit) crystal field: if the occupation is below $n\approx 1.1$, DMFT enhances the crystal field while for carrier densities $1.1\lesssim n<2$, the crystal field is renormalized in the opposite direction. As a result, the different curves in Fig.~\ref{fig:polarization_n} for different values of $U$ all intersect roughly at $n\approx 1.1$. We expect that the intersection point shifts if the explicit crystal field $\Delta$ (here set to zero) or the ratio of $J/U$ is varied but the qualitative behavior should remain the same for a range of parameter values.

\subsubsection{Low-energy properties}
\begin{figure}
\includegraphics[width=1\linewidth]{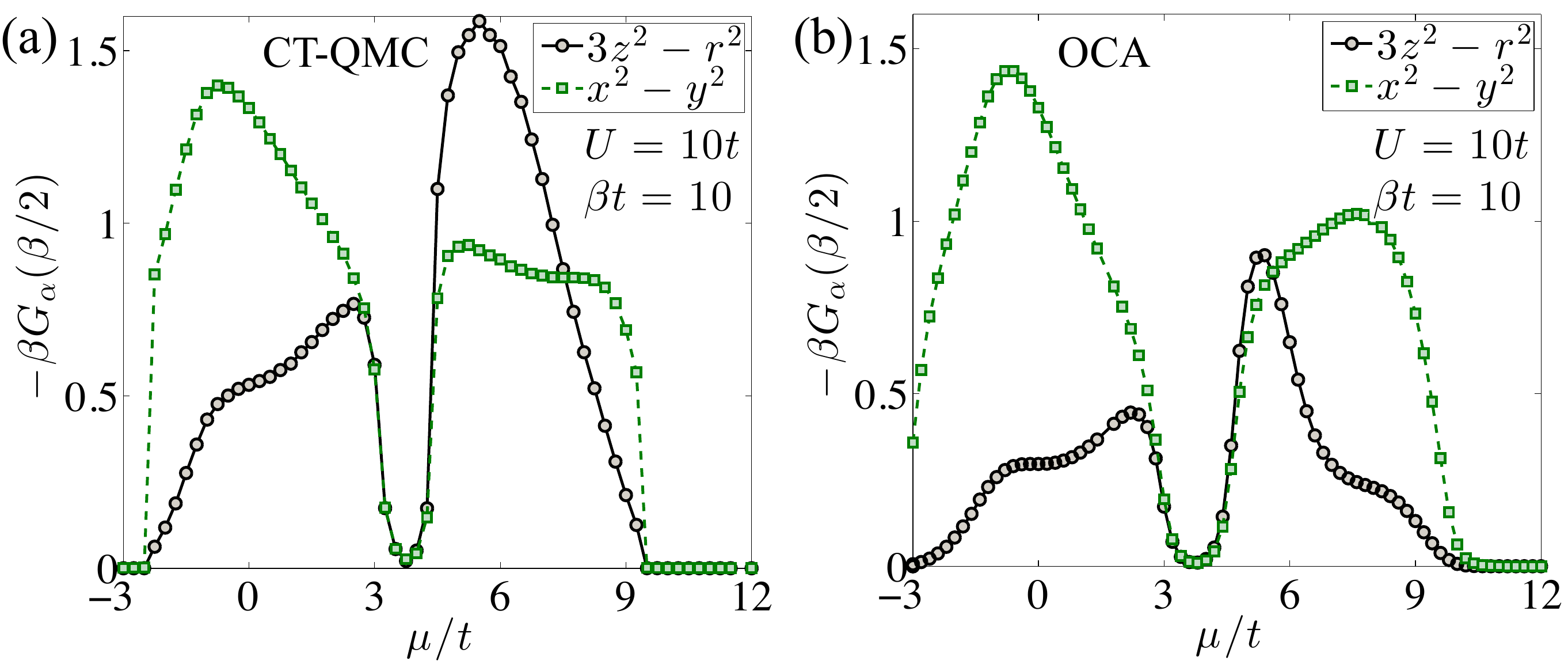}
\caption{Orbital-resolved spectral density at the chemical potential $\mu$ as function of $\mu$, see Eq.~\eqref{eq:A} within (a) DMFT(CT-QMC) and (b) DMFT(OCA). $U=10t$, $J=t$ and $\beta t=10$.}
\label{fig:A_mu}
\end{figure}

The orbital polarization discussed above is a thermodynamic quantity displaying clear signatures of correlation effects as function of carrier density. Here, we address the effects of correlations on the low-energy properties of single particle-excitations. We first consider the orbital-resolved single-particle spectral density measured at the chemical potential $\mu$, $A_{\alpha}(0)$. The exact relation\cite{Gull:2009}
\begin{equation}
-\beta G_{\alpha}(\beta/2)=\int\frac{d\omega}{2\pi T}\frac{A_{\alpha}(\omega)}{\cosh[\omega/(2T)]}
\label{eq:A}
\end{equation}
shows that $A_{\alpha}(0)$ can be estimated from the Matsubara Green's function at $\beta/2$ for low temperatures. In Eq.~\eqref{eq:A}, $\alpha=3z^2-r^2$, $x^2-y^2$ and a trace over spin-degrees is implicit. Figure~\ref{fig:A_mu} shows $-\beta G_{\alpha}(\beta/2)$ at $\beta t=10$ as function of $\mu$ for $U=10t$ within (a) CT-QMC and (b) OCA. In both cases, the transitions to the insulating phases at quarter and half filling show up as a sharp suppression of $-\beta G(\beta/2)$ when increasing the chemical potential $\mu$. Comparing CT-QMC with OCA, we find that overall the value of the $d_{x^2-y^2}$ component agrees rather well. However, the value of the $d_{3z^2-r^2}$ component is clearly underestimated within OCA. Note also that the insulating region at quarter filling appears slightly larger within OCA.

\begin{figure}
\includegraphics[width=1\linewidth]{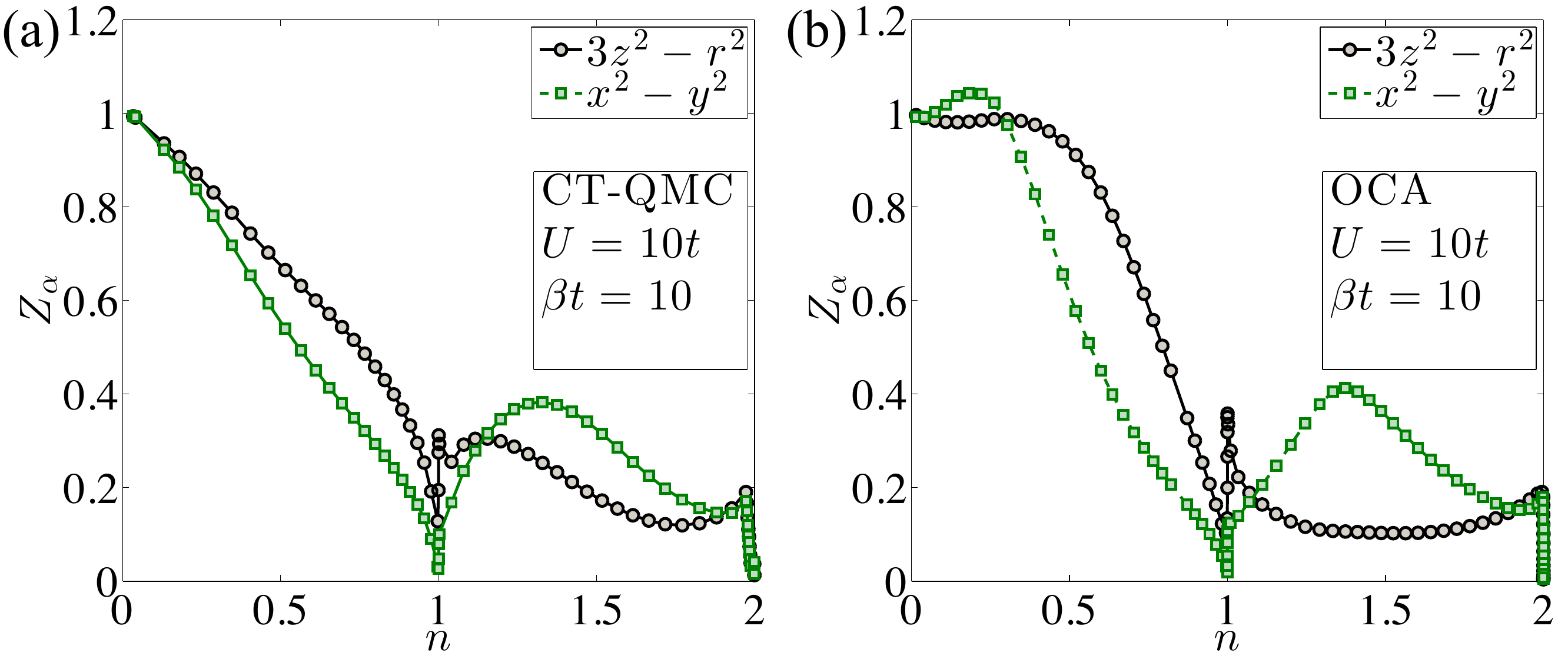}
\caption{Orbital-resolved quasiparticle weight $Z_{\alpha}$ as function of the carrier density $n$ as obtained within (a) DMFT(CT-QMC) and (b) DMFT(OCA). $Z_{\alpha}$ is not defined in the insulators at quarter filling $(n=1)$ and at half filling $(n=2)$. $U=10t$, $J=t$ and $\beta t=10$.}
\label{fig:Z_n}
\end{figure}

We next address the low-energy properties in the metallic phase. Our self-energy data are compatible with the assumption of a Fermi liquid away from the insulating phases. We therefore investigate the low-energy properties from this perspective. However, we can not rule out the existence of non-Fermi liquid phases, as observed for example in the frozen-moment phase close to the half-filled Mott insulator in a three-orbital model,\cite{Werner:2008} but its identification would require a more careful analysis of the electronic self-energy at lower temperatures which is beyond the scope of the present work. The self-energy in a Fermi liquid can be expanded for small frequencies as
\begin{equation}
\Sigma_{\alpha}(i\omega)=a_{\alpha}+ib_{\alpha}\,\omega+\mathcal{O}(\omega^2).
\end{equation}
The low-energy poles of the single-particle Green's function can then be obtained from the solution of an effective non-interacting Hamiltonian\cite{Ruegg:2008a} 
\begin{equation}
H_{\rm eff}({\bs k})=\tilde{\mathcal{E}}_{xy}({\bs k})-\tilde{\mu}+\tilde{\Delta}\hat{\tau}_z
\label{eq:Heff}
\end{equation}
Here, $\tau_z$ is the third Pauli matrix acting in orbital space and the chemical potential $\tilde{\mu}$ is chosen such that the quasiparticle density at $\tilde{\mu}$ is equal to the electron density at $\mu$. The Bloch matrix 
\begin{equation}
\tilde{\mathcal{E}}_{xy}({\bs k})=
\begin{pmatrix}
-\frac{\tilde{t}_{11}}{2}\left(\cos k_x+\cos k_y\right)&\frac{\sqrt{3}\tilde{t}_{12}}{2}\left(\cos k_x-\cos k_y\right)\\
\frac{\sqrt{3}\tilde{t}_{21}}{2}\left(\cos k_x-\cos k_y\right)&-\frac{3\tilde{t}_{22}}{2}\left(\cos k_x+\cos k_y\right)
\end{pmatrix}.
\label{eq:Htildexy}
\end{equation}
contains renormalized hopping amplitudes
\begin{equation}
\tilde{t}_{11}=Z_{1}t,\quad \tilde{t}_{22}=Z_{2}t,\quad \tilde{t}_{12}=\tilde{t}_{21}=\sqrt{Z_1Z_2}t,
\end{equation}
where the subscript $\alpha=1$ $(2)$ labels the $d_{3z^2-r^2}$ ($d_{x^2-y^2}$) orbital.
The parameters
\begin{equation}
Z_{\alpha}=\lim_{T\rightarrow 0}\left[1-\frac{{\rm Im}\ \Sigma_{\alpha}(i\omega_0)}{\omega_0}\right]^{-1},
\label{eq:Z}
\end{equation}
are the orbital-resolved quasiparticle weights. In the non-interacting limit, $Z_{\alpha}=1$ and the suppression of $Z_{\alpha}$ is a measure of how strongly correlated the metallic state is. Figure~\ref{fig:Z_n} shows the result obtained by using Eq.~\eqref{eq:Z} for finite temperatures $\beta t=10$ evaluated again within (a) CT-QMC and (b) OCA. The usage of Eq.~\eqref{eq:Z} at finite temperatures should be taken with some care, because it implicitly assumes the existence of well-defined quasiparticles and a reliable extrapolation to zero-temperatures. But in any case, it provides a useful characterization of the low-energy properties of the self-energy, thereby allowing us to contrast CT-QMC to OCA for various carrier densities. Note that in the insulating phases at $n=1$ and $n=2$, $Z_{\alpha}$ is not defined. The first-order nature of the metal-insulator transition is manifest by the fact that $Z_{\alpha}$ jumps across the insulating phases. We also see that the $d_{x^2-y^2}$ orbital is more strongly correlated than the $d_{3z^2-r^2}$ orbital for $0<n<1$ and vice-versa for $1<n<2$. Overall, we find that the OCA prediction for the quasiparticle weight of the $d_{x^2-y^2}$ orbital is more accurate than the one for the $d_{3z^2-r^2}$ orbital. This is similar to the observation made for the spectral weight at $\mu$ discussed above in Fig.~\ref{fig:A_mu}.
 
\begin{figure}
\includegraphics[width=1\linewidth]{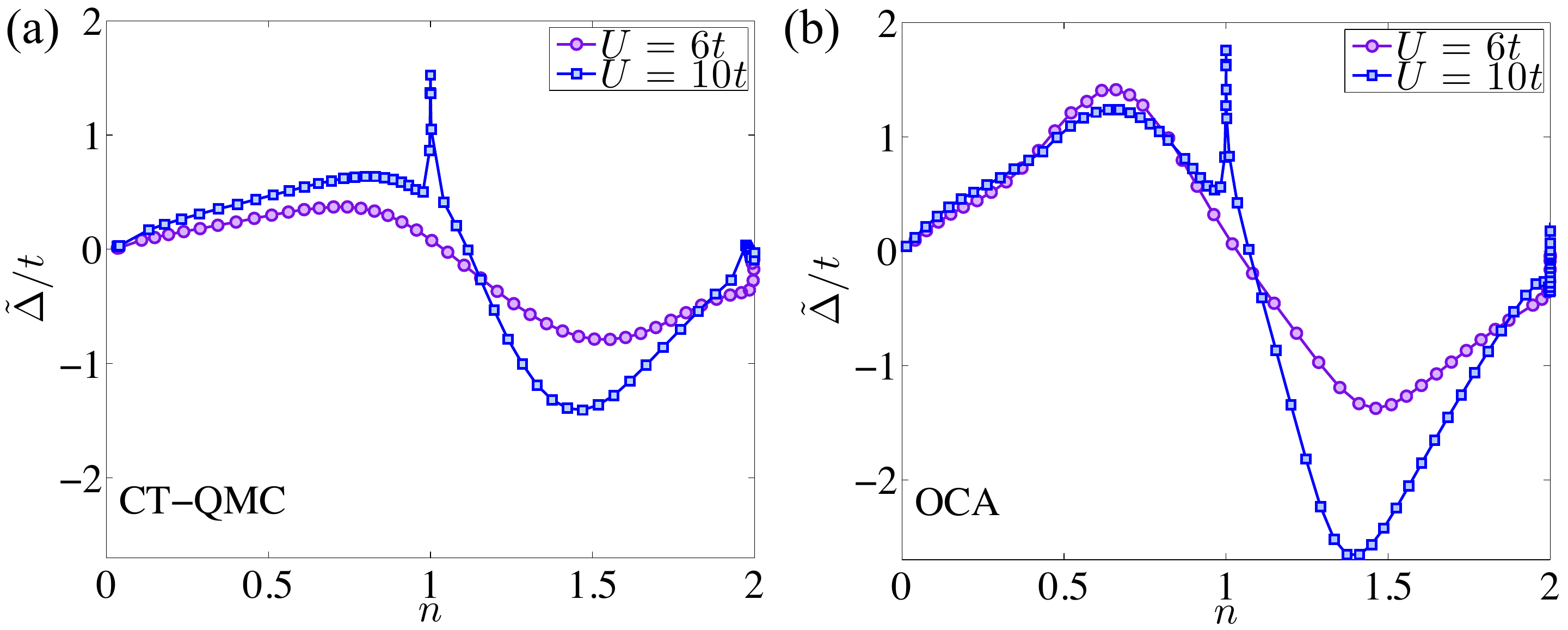}
\caption{The dependence of the effective crystal-field splitting $\tilde{\Delta}$ on the electron density $n$ in the monolayer model ($L=1$) for $\beta t=10$ within (a) DMFT(CT-QMC) and (b) DMFT(OCA). Shown are results for various interaction strength $U=6t$ and $U=10t$ at fixed Hund's coupling $J=t$.}
\label{fig:eff_CF}
\end{figure}

A further important quantity entering the effective Hamiltonian Eq.~\eqref{eq:Heff} is the effective crystal field $\tilde{\Delta}$ which is defined as
\begin{equation}
\tilde{\Delta}=\lim_{T\rightarrow 0}\sum_{\alpha}Z_{\alpha}{\rm Re}\left[\Sigma_{\alpha}(i\omega_0)\right].
\label{eq:Deltaeff}
\end{equation}
A positive value of $\tilde{\Delta}$ suppresses occupation of the $d_{3z^2-r^2}$ orbital and, thus, favors a negative orbital polarization ($p<0$).
At finite temperatures, we estimate $\tilde{\Delta}$ by extrapolating $\Sigma_{\alpha}(i\omega)$ to $\omega\rightarrow 0$. The results are shown in Fig.~\ref{fig:eff_CF} for an intermediate ($U=6t$) and a large ($U=10t$) value of the interaction. For both values, $\tilde{\Delta}$ changes sign from positive to negative when increasing the carrier density $n$. ($\tilde{\Delta}$ is not defined in the insulating phases.) Importantly, $\tilde{\Delta}<0$ favors a positive orbital polarization in the vicinity of $n=1.5$, which is consistent with the discussion of the orbital polarization in Sec.~\ref{sec:OP}. 

\section{Conclusions}
\label{sec:conclusions}
In summary, we used the OCA, ED and CT-QMC impurity solvers within the layer-DMFT framework to investigate the $e_g$-Hubbard model in a thin film geometry relevant for sandwich structures involving rare earth nickelates. The advantage of OCA over CT-QMC/ED is its relatively low numerical cost (approximatively 10\% of CT-QMC at the presented temperatures). Moreover, we find that the OCA is accurate (and efficient, i.e.~both the OCA and DMFT self-consistency converges rapidly) in the insulating regime. In contrast, the accuracy in the metallic regime is reduced and only qualitative agreement can be expected. Interestingly, this holds even for large interactions when entering the metallic regime by tuning the carrier density away from commensurate values. On the other hand, the ED solver gives results consistent with CT-QMC for arbitrary interaction strength, but for the fixed small fictive temperature, we found better convergence in the metallic regime. An issue we observed within ED is that the convergence with respect to the number of bath sites can be slow in the considered situations, i.e.~for layered multi-orbital systems.

From the physical point-of-view, we presented several interesting aspects of the $e_g$-Hubbard model in thin film geometries. First, we discussed the thickness dependence of the metal-insulator transition and the orbital polarization. As expected and consistent with experiments,\cite{Boris:2011,Liu:2011} we found that the metallic phase is suppressed for ultra thin films. We also demonstrated that orbital polarization of the {\em unstrained} system ($\Delta=0$) is a surface phenomena which quickly dies off about three layers away from the interface. Therefore, an average polarization that decreases roughly as the inverse of the film width for films thicker than 3 atomic layers is expected in the unstrained situation. This should be readily observable in experiments, using e.g. the x-ray linear dichroism (XLD) as in Ref.~\onlinecite{Chakhalian:2011}, which averages the Ni signal over the entire width of the thin film. Deviations from this trend would indicate that also the inner layer contribute to the signal, which points towards a strain effect, i.e.~an orbital polarization induced by a crystal field $\Delta\neq 0$. Moreover, Refs.~\onlinecite{Benckiser:2011,Wu:2013} demonstrated that soft x-ray reflectivity can resolve as little as a 3\% difference in orbital polarization between layers in LaNiO$_3$. Hence, this technique is able to produce spatially resolved data that can be quantitatively compared with the trends predicted in our layer-resolved calculations. Interestingly, the data presented in Ref.~\onlinecite{Wu:2013} do indicate an enhanced orbital polarization of the boundary layers.

Second, we investigated how physical quantities depend on the carrier density and the interaction strength for the monolayer system. Interestingly, the orbital polarization depends quite strongly on both the carrier density and the interaction strength. This dependence is most prominent in the metallic phase for electron densities in the range $1<n<2$ where we observe a correlation-induced sign change of the orbital polarization. Our results are also interesting in view of the important problem of identifying the correct effective model for nickelate heterostructures. Several recent studies\cite{Wang:2012,Parragh:2013} revealed notable differences between an ``$e_g$-only" and a $d-p$ model, which explicitly treats the hybridization with (uncorrelated) oxygen $p$-orbitals. One of the effects of including the $p$-bands in the low-energy model is that the occupation of the $d$-manifold gets closer to half-filling. This makes the effect of the Hund coupling stronger than in the quarter-filled $e_g$-only model, affecting, for example, the value of the orbital polarization. Our results confirm this observation by demonstrating a similar effect in the $e_g$-only model upon changing the carrier concentration.

\begin{acknowledgements}
We thank Andy Millis for previous collaboration and many stimulating discussions. We thank Philipp Hansmann for providing the HF-QMC data of Ref.~\onlinecite{Hansmann:2010} and Ansgar Liebsch for a helpful correspondence. H.H.H. sincerely thanks precious suggestions and helpful discussions from Chungwei Lin and Dominika Zgid. A.R., H.H.H and G.A.F. acknowledge financial support through ARO Grants No.~W911NF-09-1-0527 and W911NF-12-1-0573, NSF Grant No.~DMR-0955778 and DARPA grant D13AP00052. A.R. acknowledge partial support from the Swiss National Science Foundation. The DMFT(OCA) calculations were performed at the Max Plank Institute for the Physics of Complex Systems in Dresden and the DMFT(CT-QMC) on the Brutus cluster at ETHZ. The authors acknowledge the Texas Advanced Computing Center (TACC) at the University of Texas at Austin for providing some of the computational resources used in this work (http://www.tacc.utexas.edu).
\end{acknowledgements}

\begin{widetext}
\appendix
\section{Dependence on the number of bath sites in the DMFT(ED) scheme}

\begin{figure*}[ht]
\includegraphics[width=0.49\linewidth]{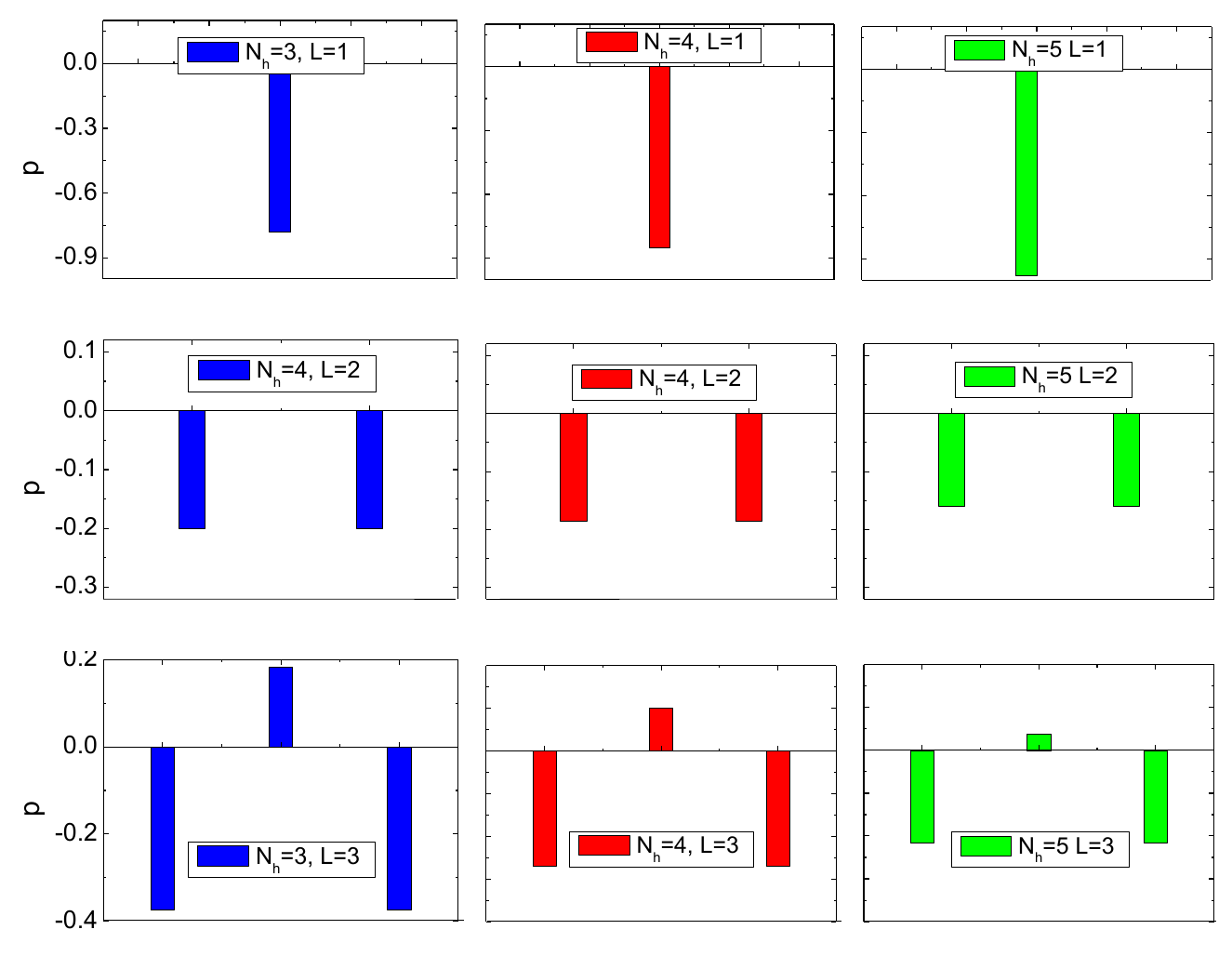}
\includegraphics[width=0.49\linewidth]{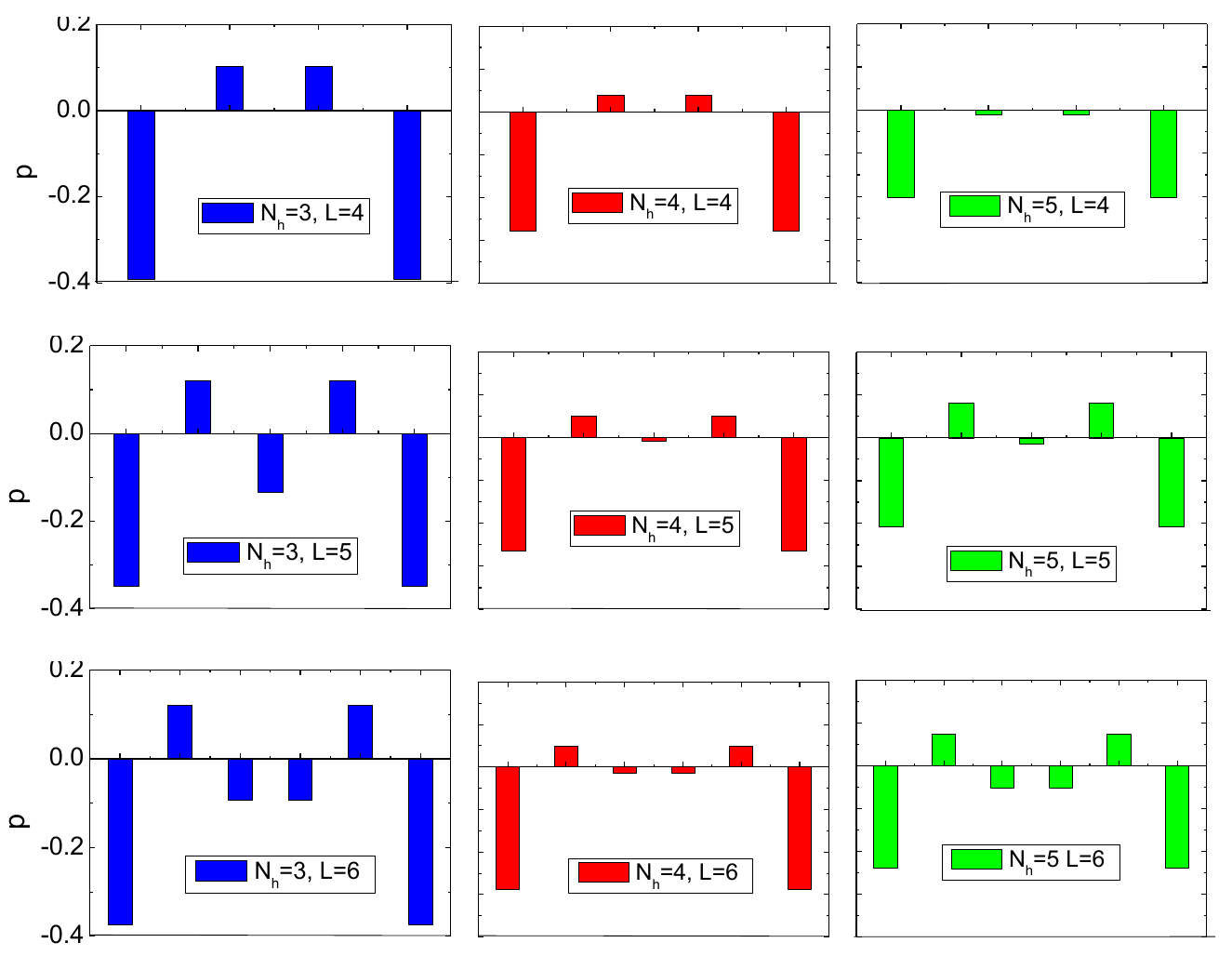}
\caption{Summary of the layer-resolved orbital polarization within DMFT(ED) for different number of bath sites $N_h=3$, $4$ and 5 at $U=12t$ and a fictive temperature $T_0=0.0005t$.}
\label{fig:OP_N}
\end{figure*}

The number of bath sites within DMFT(ED) is an important parameter. All the results presented in the main text are obtained using $N_h=5$ bath sites per orbital. However, we also performed calculations for fewer bath sites $N_h=3$ and 4. In Fig.~\ref{fig:OP_N}, in order to provide an indication on the convergence of the ED results with respect to $N_h$, we provide a summary of the layer-resolved orbital polarization for $N_h=3$, 4 and 5. Ideally, to have accurate results, the ED for the largest $N_h$ available is close to the limit $N_h\rightarrow\infty$. From the observed behavior of the orbital polarization, we conclude that the convergence with $N_h$ is relatively slow for the considered systems.

\end{widetext}

\bibliography{biblio}

\end{document}